\documentclass[twocolumn,trackchanges,twocolappendix]{aastex631}

\usepackage{bm}

\received{}
\revised{}
\accepted{}
\submitjournal{ApJ}

\shorttitle{Turbulent reacceleration and mega halo}
\shortauthors{Nishiwaki et al.}

\begin{document}


\title{Efficiency of turbulent reacceleration by solenoidal turbulence and its application to the origin of radio mega halos in cluster outskirts}

\author[0000-0003-2370-0475]{Kosuke Nishiwaki}
\affiliation{Institute for Cosmic Ray Research, The University of Tokyo, 5-1-5 Kashiwanoha, Kashiwa, Chiba 277-8582, Japan}

\author[0000-0003-4195-8613]{Gianfranco Brunetti}
\affiliation{INAF Istituto di Radioastronomia, Via P.Gobetti 101, I-40129 Bologna, Italy}

\author[0000-0002-2821-7928]{Franco Vazza}
\affiliation{Dipartimento di Fisica e Astronomia, Universita di Bologna, Via Gobetti 93/2, I-40129 Bologna, Italy}
\affiliation{INAF Istituto di Radioastronomia, Via P.Gobetti 101, I-40129 Bologna, Italy}
\affiliation{Hamburger Sternwarte, Universit\"{a}t Hamburg, Gojenbergsweg 112, 41029 Hamburg, Germany}

\author[0000-0003-1063-3541]{Claudio Gheller}
\affiliation{INAF Istituto di Radioastronomia, Via P.Gobetti 101, I-40129 Bologna, Italy}

\begin{abstract}
Recent radio observations with Low-Frequency Array (LOFAR) discovered diffuse emission extending beyond the scale of classical radio halos.
The presence of such mega halos indicates that the amplification of the magnetic field and acceleration of relativistic particles are working in the cluster outskirts, presumably due to the combination of shocks and turbulence that dissipate energy in these regions.
Cosmological magnetohydrodynamical (MHD) simulations of galaxy clusters suggest that solenoidal turbulence has a significant energy budget in the outskirts of galaxy clusters.
In this paper, we explore the possibility that this turbulence contributes to the emission observed in mega halos through second-order Fermi acceleration of relativistic particles and the magnetic field amplification by the dynamo.
We focus on the case of Abell 2255 and find that this scenario can explain the basic properties of the diffuse emission component that is observed under assumptions that are used in previous literature.
More specifically, we conduct a numerical follow-up, solving the Fokker--Planck equation using a snapshot of a MHD simulation and deducing the synchrotron brightness integrated along the lines of sight.
We find that a volume-filling emission, ranging between 30 and almost 100\% of the projected area depending on our assumptions on the particle diffusion and transport, can be detected at LOFAR sensitivities.
Assuming a magnetic field $B\sim0.2\mu$G, as derived from a dynamo model applied to the emitting region, we find that the observed brightness can be matched when $\sim$1\% level of the solenoidal turbulent energy flux is channeled into particle acceleration.
\end{abstract}

\keywords{Galaxy clusters (584)}

\section{Introduction}\label{sec:intro}
Galaxy clusters are filled with a hot intra-cluster medium (ICM), that has a characteristic temperature similar to the the cluster's virial temperature.
This suggests that the ICM is heated by the gravitational energy released in the hierarchical merger and accretion processes of clusters \citep[e.g.,][]{Press_Schechter_1974,Kravtsov_Borgani_2012}.
A fraction of the energy can also be channeled into non-thermal components, such as relativistic particles and magnetic fields.
Shocks and turbulence could be favorable sites for the particle acceleration and the amplification of the field \citep[see][for review]{Brunetti_Jones_review}.
\par

Radio observations of galaxy clusters probe those non-thermal components by studying diffuse synchrotron emission of relativistic (cosmic-ray) electrons (CRe).
Radio halo is a diffuse emission with an extension of $\sim$1 Mpc, often found in the central region of merging clusters \citep[see][for review]{vanWeeren_review}. 
The radiative cooling time of CRe is significantly shorter than the time required for diffusion or advection over $\sim 1~{\rm Mpc}$, implying that there is an {\it in situ} mechanism that produces CRe.
Reacceleration by merger-induced turbulence is the most plausible scenario \citep[e.g.,][]{Brunetti_2001MNRAS.320..365B,Petrosian_2001,Fujita_2003ApJ...584..190F,Cassano_Brunetti_2005}, although cosmic-ray protons (CRp) in the ICM may be important ingredients in the physics of those phenomena \citep[e.g.,][]{Dennison_1980,Blasi_1999APh....12..169B}.
For example, they can provide seed CRe to re-accelerate through the hadronic {\it pp} collision with the thermal protons in the ICM \citep[e.g.,][]{Brunetti_Lazarian_2011MNRAS.410..127B,Brunetti_2017MNRAS.472.1506B,Pinzke_2017,Nishiwaki_2022ApJ}. 
It has been shown that the observed statistical properties of radio halos are in line with the reacceleration model considering the resonant interaction between compressible turbulence \citep[][]{Cassano_Brunetti_2005,Nishiwaki_2022ApJ,Cassano_2023}.
Reacceleration by turbulence is also proposed to explain radio emission detected on larger scales, such as radio bridges \citep[][]{BV20} that are radio filaments connecting massive pairs in the early stage of mergers discovered by the Low Frequency Array (LOFAR) \citep[][]{Govoni_2019,Botteon_2020}.


More recently, \citet{Cuciti_Megahalo} reported the existence of radio ``mega halos" in four clusters, using the LOFAR observation.
The volume of the mega halos is almost 30 times larger than that of radio halos, suggesting that the entire volume of the cluster is filled with CRe and magnetic field.
The radio power of mega halos is comparable to or even larger than that of classical halos.
The detection of synchrotron radiation at the large distance (1 - 2 Mpc) from the cluster center also constrains the magnetic field strength in this region.
Since the pressure of non-thermal components, including magnetic field and CRe, should be smaller than the thermal one as indicated from the observations and numerical simulations \citep[e.g.,][]{Vazza_2016,Eckert_2019}, the magnetic field should be in the range of $0.1\mu{\rm G} \lesssim B \lesssim 1.7\mu{\rm G}$ \citep[][]{Botteon_A2255}.
The lower bound ($0.1\mu{\rm G}$) is at least one order of magnitude larger than the value expected by the compression of primordial fields.
One possible mechanism of this non-linear amplification of the field is dynamo in a turbulent medium. 
\par


Cosmological simulations of galaxy clusters suggest that turbulence and shocks driven by continuous accretion of matter fill the entire volume of the cluster up to the virial radius \citep[e.g.,][]{Vazza_2011A&A...529A..17V,Nelson_2014,Miniati_2015ApJ...800...60M,Steinwandel_2023}.
As in the cluster center, the ICM in the outskirts is a weakly-collisional plasma, and the perturbations would cause instabilities that effectively reduce the mean free path (mfp) of thermal protons \citep[e.g.,][]{Schekochihin_2005ApJ...629..139S,Kunz_2011MNRAS.410.2446K,Brunetti_Lazarian_2011}, potentially establishing a well-developed inertial range \citep[e.g.,][]{Schekochihin_2009ApJS..182..310S}.

Indeed, the infall of the clumps of mass drives the turbulence with a typical scale of a few 100 kpc in the cluster outskirts \citep[e.g.][]{Vazza_2017MNRAS.464..210V}. The timescale of the turbulent cascade, $t_{\rm cas}\sim 600$ Myr \citep[e.g.,][]{Brunetti_Lazarian_2007}, is much shorter than the Hubble time, which allows the turbulent dynamo to work in that region.

\par




Nowadays, the best studied case of the cluster hosing diffuse radio emission on the entire cluster volume is the case of Abell 2255 (hereafter A2255) \citep[][]{Botteon_A2255}.
A2255 is a nearby ($z = 0.0806$) cluster, which shows a complex dynamical state in optical and X-ray observations \citep[e.g.,][]{Burns_1995ApJ...446..583B,Yuan_2003,Golovich_2019,Feretti_1997,Akamatsu_2017}. 
The cluster is known to host a radio halo and radio relics, and they have been studied in wide range of frequencies \citep[e.g.,][]{Jaffe_1979,Feretti_1997,Govoni_2005,Pizzo_2008,Pizzo_2009,Botteon_2020_A2255}.

More recently, LOFAR observations found that the cluster hosts diffuse emission extending in very large scales and enveloping the classical halo and relics \citep[][]{Botteon_A2255}.
Such emission is complex, showing a number of relic-like features embedded in a truly diffuse component, and a spectral index distribution between 40 - 144 MHz ranging from 0.6 to 2.5.
The flat spectrum emission is coincident with the radio relics located in the north and southwest parts of the cluster, while the steep emission is associated with the diffuse component.
\par

In this paper, we attempt to explore the possibility that turbulence contributes to the observed emission via magnetic field amplification and particle reacceleration.
We use cosmological MHD simulation of a cluster to examine the turbulent and magnetic fields in the cluster outskirts.
In Sect.~\ref{sec:simulation}, we describe the setup of the MHD simulation.
In Sect.~\ref{sec:theory}, we review the reacceleration model and claim that that is compatible with the observed spectrum of the mega halo.
In Sect.~\ref{sec:FP}, we numerically solve the Fokker--Planck (FP) equation, considering the distribution of turbulence and the projection along the line of sight.
The limitations of our models are discussed in Sect.~\ref{sec:limitation}.
Finally, we summarize the results in Sect.~\ref{sec:concl}.

\section{Cosmological MHD simulation}\label{sec:simulation}

To examine the property of the ICM and turbulence in cluster outskirts, we use a snapshot of a high-resolution cosmological ideal MHD simulation of a massive galaxy cluster, produced with grid code ENZO \citep[][]{Enzo2014}.
We use the same simulated cluster as in \citet{Botteon_A2255}, which has a mass of $M_{200}=8.65 \times  10^{14}~ \rm M_{\odot}$ at $z=0$.
This simulation includes eight levels of Adaptive Mesh Refinement (AMR) to increase the spatial and force resolution within the virial volume of the cluster, reaching a peak spatial resolution of $\Delta x = 3.95 \rm ~ kpc/cell$ (comoving) \citep[][]{vazza18dynamo}.
\par

The simulation was started assuming a uniform  "primordial" seed magnetic field of $B_0=0.1 \rm ~nG$ (comoving) at $z=40$. 
The low redshift properties of the magnetic field in the cluster volume are found to be fairly independent of the exact origin scenario, due to the effect of efficient small-scale dynamo amplification \citep[][]{vazza18dynamo}.
At the late stage of the cluster evolution, most of the central volume within $<1$ Mpc is simulated with the finest resolution $\Delta x = 3.95 \rm ~ kpc/cell$.
The virial volume of clusters is refined at least up to $\Delta x= 15.8 \rm ~ kpc/cell$ \citep[][]{Dominguez-Fernandez_2019MNRAS.486..623D}.
At 1-2 Mpc from the center, the spatial resolution is comparable to, or coarser than, the MHD scale, $l_{\rm A}\sim 1$ kpc, where the turbulent velocity becomes comparable to the Alfv\'en velocity, so the field amplification may be underestimated in the simulation.
Thus, we evaluate the field strength in this region in a post-process (see Sect.~\ref{sec:theory} for the detail).
\par

The turbulent kinetic flux is calculated with small-scale filtering explained in \citet{Vazza_2017MNRAS.464..210V}, which allows us to reconstruct (and remove) the contribution from shock waves to the total turbulent energy budget. 
For each cell of the simulation, the dispersion of the velocity field $\sigma_{v}(L)$ is measured within a scale $L$.
The outer scale of the turbulence $\Lambda$ is defined as the scale where the change in $\sigma_{v}(L)$ with increasing $L$ becomes sufficiently small \citep[see][for the detail]{Vazza_2017MNRAS.464..210V}.
The turbulent kinetic energy flux in each simulated cell can be calculated as:

\begin{equation}
    \mathcal{F}_{\rm turb} =  \frac{1}{2} \frac{\rho \sigma_{v}^3(L)}{L} (\Delta x)^3\:,
    \label{eq:Fsol}
\end{equation}

\noindent
where $\rho$ is the gas mass density in the cell and $\Delta x$ is the cell size.
Based on previous works, we assume that the turbulent spectrum in the inertial range roughly follows the Kolmogorov scaling \citep[e.g.][]{Vazza_2011A&A...529A..17V,Vazza_2017MNRAS.464..210V}, and so the value of $\mathcal{F}_{\rm turb}$ is insensitive to the specific value of $L$ as long as it is in the inertial range.
To study the quantities in the cluster outskirts, we extract a box of a region of 1.6 Mpc on a side located at 1.2 Mpc from the center of the simulated cluster.
Each cell has a volume of $16^3~{\rm kpc}^3$, and the box is composed of $10^6$ cells.
In this region, the outer scale $\Lambda$ is found to be typically $\Lambda \approx 200~{\rm kpc}$.
In the following, we use the turbulent velocity measured at $L = 160~{\rm kpc}$.

\begin{figure*}
    \plotone{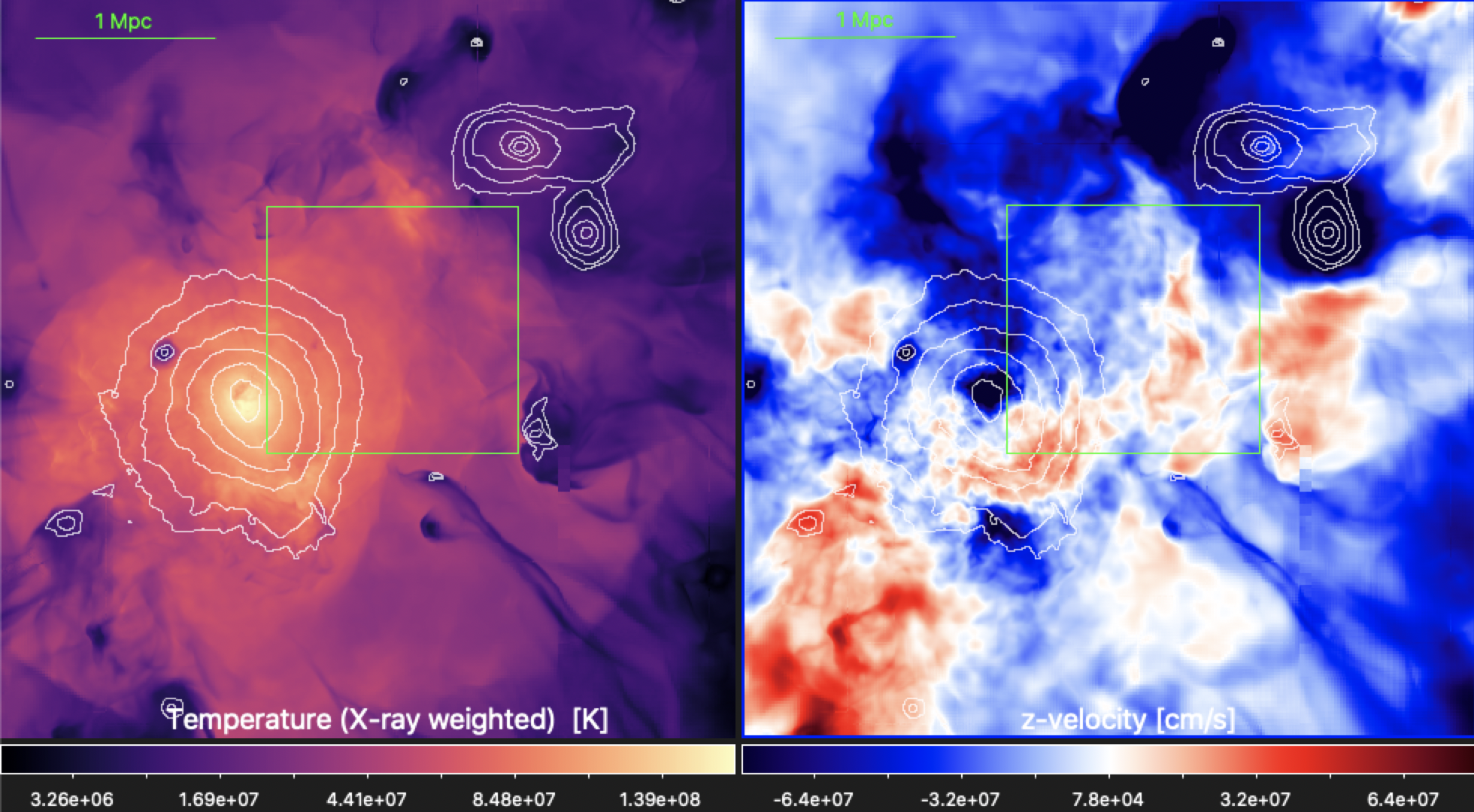}
    \caption{Maps of the X-ray weighted gas temperature along the line of sight (left) and of the gas velocity along the line of sight (right) for the simulation snapshot analyzed in this work. The additional white contours show the regions which can be approximately detected with X-observations in the 0.5-2 keV energy band, while the green square shows the location of the box used to extract the turbulent flow properties used in Sec.\ref{sec:FP}. Each image is made of 1024 $\times$ 1024 pixels.}
    \label{fig:sim}
\end{figure*}

Fig. \ref{fig:sim} gives the visual impression of the line of sight gas velocity and of the gas temperature in the volume of the simulated clusters, in particular in the box used to extract the physical parameters used in our Fokker--Planck calculations.
The selected box is at the cluster periphery, in a region only partially detectable through typical X-ray observations, and along the direction of a prominent filamentary accretion, but without the presence of massive substructures, or prominent shock waves.
\par

\begin{figure*}
    \centering
    \plottwo{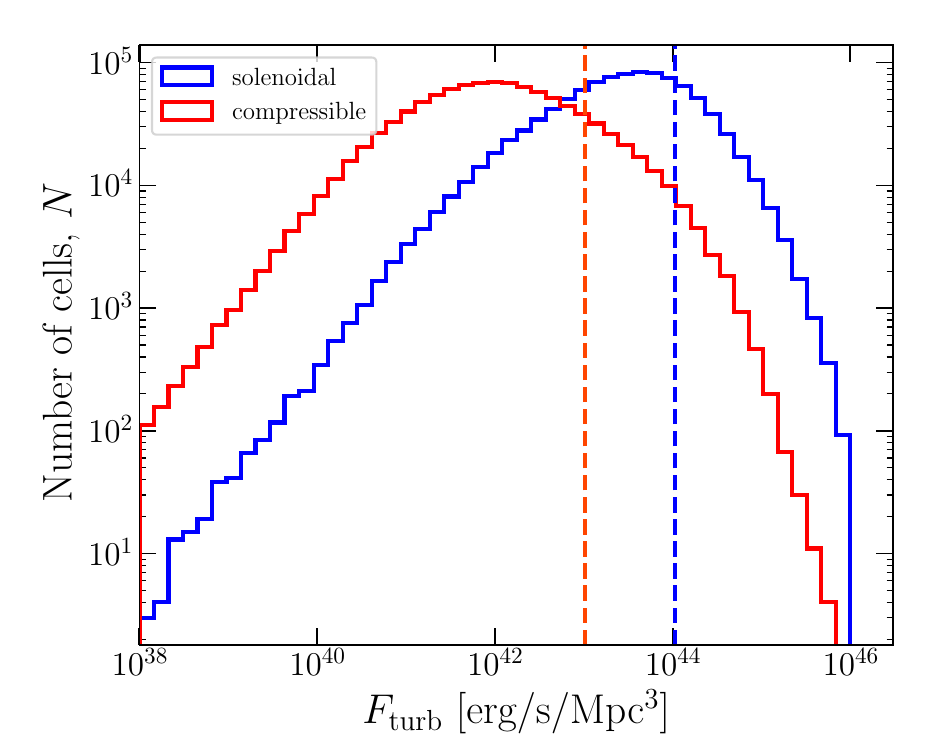}{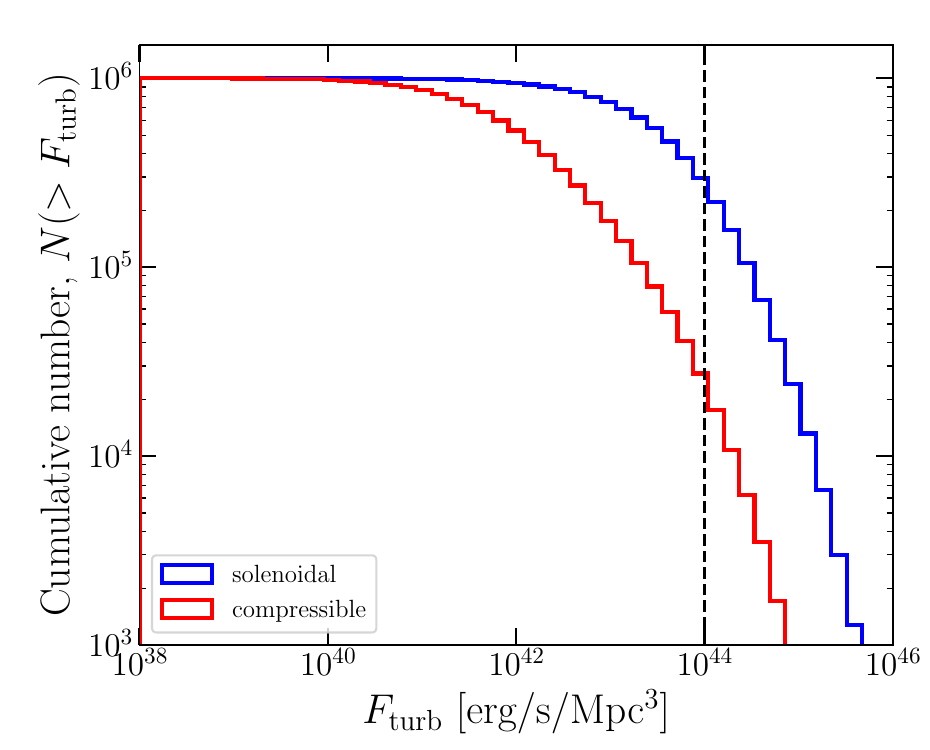}
    \caption{Left: Histogram of turbulent energy flux per unit volume for solenoidal (blue) and compressible (red) modes in the cluster peripheral region shown in Fig.~\ref{fig:sim}.
    The vertical axis shows the number of cells in the simulated box.
    The dotted vertical lines show the mean values for each mode. The total number of cells in the extracted box is $N=10^6$.
    Right: Cumulative number of cells with the turbulent kinetic flux larger than $F_{\rm turb}$. The vertical dashed line shows $F_{\rm turb} = 10^{44}~{\rm erg/s/Mpc^3}$ }
    \label{fig:F_turb}
\end{figure*}

Fig.~\ref{fig:F_turb} (left panel) shows the histogram of turbulent kinetic flux in the extracted region.
We decompose the turbulent velocity into solenoidal ($\nabla\cdot\bm{v} = 0$) and compressible ($\nabla \times \bm{v} = 0$) modes, using the procedure of  \citet{Vazza_2017MNRAS.464..210V}.
We find that the solenoidal mode typically has a larger kinetic flux than the compressible mode, in line with previous simulations \citep[e.g.,][]{Miniati_2015ApJ...800...60M,Porter_2015ApJ...810...93P,Vazza_2017MNRAS.464..210V}. 
The mean value of the solenoidal kinetic flux per unit volume is $F_{\rm turb} = 1.1\times10^{44}~{\rm erg/s/Mpc^3}$, which is almost 10 times larger than the compressible component.
In the right panel, we show the cumulative number of cells that have turbulent kinetic flux larger than $F_{\rm turb}$. More than 30\% of the cell has the solenoidal turbulent flux larger than $10^{44}~{\rm erg/s/Mpc^3}$, while the fraction decreases to $\approx3$\% in the case of the compressible mode. 
Note that we are considering the sector where the feature of mass accretion from the nearby cluster can be seen (Fig.~\ref{fig:sim}) and it is more turbulent than other sectors of cluster outskirts (see also Sect.~\ref{sec:limitation}). 
For comparison, we study other regions with the same volume and the distance from the cluster center and find a factor $\sim$5 smaller turbulent flux, whereas the solenoidal mode dominates the compressible one in every region.
In the following sections, we focus on the solenoidal velocity to calculate the acceleration efficiency and the dynamo field. Note that $F_{\rm turb}$ denotes only the volumetric turbulent kinetic flux of the solenoidal component.


\par

\section{CR reacceleration and field amplification by solenoidal turbulence}\label{sec:theory}

Turbulence driven through the formation process of galaxy clusters is typically sub-sonic ($M_{\rm s} < 1$) and super-Alfv\'enic ($M_{\rm A} > 1$) \citep[][]{Brunetti_Lazarian_2007}.
A fraction of the turbulent energy can be converted into non-thermal components such as cosmic rays and the magnetic field through stochastic acceleration and turbulent dynamo.
Turbulence in astrophysical environments may accelerate particles through different mechanisms, including resonant and non-resonant mechanisms \citep[e.g.,][]{Ptuskin_1988,Schlickeiser_Miller_1998,Cho_Lazarian_2006,Brunetti_Lazarian_2007,Lynn_2014,BL16,Bustard_Oh_2022,Lemoine_2021,Lazarian_Xu_2023}.
Recent MHD simulations of galaxy clusters suggested that the turbulence in the ICM is dominated by the solenoidal (incompressible) mode \citep[e.g.,][]{Miniati_2015ApJ...800...60M,Vazza_2017MNRAS.464..210V}.
One possible acceleration mechanism working in incompressible turbulence is the acceleration due to the interaction between magnetic field and particles diffusing in super-Alfv\'enic turbulence \citep[][]{BL16,BV20}.

\par

In turbulent reconnection theory \citep[][]{LV99}, the Alfv\'en scale $l_{\rm A} \equiv LM_{\rm A}^{-3}$ (for the Kolmogorov scaling) is the dominant scale, where the reconnection speed may reach $v_{\rm rec} \sim v_{\rm A}$. 
In the regions where the magnetic field dissipates due to the reconnection, the particles trapped in contracting islands gain energy through a mechanism that is similar to the first-order Fermi mechanism \citep[][]{Kowal_2012PhRvL,delValle_2016}. 
On the contrary, the particles are expected to cool in the regions where the dynamo is efficient and the magnetic field lines diverge.

\par

According to \citet{BL16} particles diffusing through this complex pattern experience a second-order acceleration mechanism. 
In this mechanism, the diffusion coefficient in the momentum space is 
\begin{equation}\label{eq:Dpp}
D_{pp} = 3\sqrt{\frac{5}{6}}\frac{c_s^2}{c}\frac{\sqrt{\beta_{\rm pl}}}{L}M_s^3\psi^{-3}p^2,
\end{equation}
where $p$ is the momentum of the particle, $\beta_{\rm pl}\equiv 2\gamma_{\rm ad}^{-1}c_{\rm s}^2/v_{\rm A}^2$ is the plasma beta with $\gamma_{\rm ad} = 5/3$ is the adiabatic index, and $\psi \equiv \lambda_{\rm mfp}/l_{\rm A}$ with $\lambda_{\rm mfp}$ is the mean free path (mfp) of the particle.
The mfp is an important parameter in the model as it determines the spatial diffusion coefficient,
\begin{equation} \label{eq:D_diff}
    D\sim\frac{1}{3}c\lambda_{\rm mfp} \sim 10^{31} {\rm cm}^2{\rm s}^{-1}\left(\frac{\psi}{0.5}\right)\left(\frac{l_{\rm A}}{1~{\rm kpc}}\right),
\end{equation}
and consequently the efficiency of particle crossing regions of the size $l_{\rm A}$. 
Combining the requirement that the fractional change of momentum in each scattering should be $\Delta p/p \ll 1$ and the effect of particle scattering due to mirroring in a super-Alfv\'enic flow, $\psi$ should satisfy $0.01 \ll \psi \lesssim 0.5$; $\psi\sim 0.5$ is the reference value that has been motivated and used in the previous studies \citep[][]{BL16,BV20}.

We consider that the solenoidal turbulence with super-Alfv\'enic injection velocity shows a power spectrum with a well-established inertial range.
In such a situation, a fixed fraction ($\eta_B\approx0.05$) of the turbulent energy flux is consumed through the amplification of the magnetic field \citep[e.g.,][]{Cho_2009ApJ...693.1449C,Beresnyak_2012,Xu_Lazarian_2016}.
MHD simulations of galaxy clusters succeed in resolving the small-scale turbulent dynamo in the central region \citep[e.g.,][]{ZuHone_2011ApJ...743...16Z,vazza18dynamo,Dominguez-Fernandez_2019MNRAS.486..623D,Steinwandel_2022}, but this effect is quenched in the cluster periphery due to the limited resolution.
Thus, we adopt the same procedure as \citet{BV20} and estimate the dynamo-amplified field in post-processing, in the output of the MHD simulation (Sect.~\ref{sec:simulation}). 
Assuming that a fraction $\eta_B$ of the turbulent kinetic energy is channeled into the field amplification, the field strength can be estimated as 
\begin{equation}\label{eq:dynamo}
\frac{B^2}{8\pi} \sim \eta_{B}F_{\rm turb}t_{\rm eddy} \sim  \frac{1}{2}\eta_B\rho_{\rm ICM}\delta v_{\rm turb}^2,
\end{equation}
where $t_{\rm eddy} = L/\delta v_{\rm turb}$ is the eddy turn-over time.
In this case, $M_{\rm A}$ and $\eta_{B}$ are related as $M_{\rm A} \sim \eta_B^{-1/2}$.
In the simulations of galaxy clusters, the value of $\eta_B$ is as small as $\eta_B\approx 0.05$ \citep[][]{Beresnyak_Miniati_2016ApJ...817..127B,vazza18dynamo}. 
Using $\eta_{B}=0.05$, we obtain the mean field strength of $\langle B\rangle \sim0.24\mu{\rm G}$ in the simulated box, which is almost 2 times larger than the mean value of the field strength directly measured in the simulation ($\langle B_{\rm sim}\rangle = 0.11\mu$G).
The plasma beta is $\beta_{\rm pl}\approx 200-500$, which is slightly larger than that in the cluster center \citep[e.g.,][]{Brunetti_Lazarian_2007}.
Adopting the dynamo field (Eq.~(\ref{eq:dynamo})), $D_{pp}$ (Eq.~(\ref{eq:Dpp})) can be expressed as a function of $\eta_{B}$ as \citep[][]{BV20}
\begin{equation}\label{eq:Dpp_dynamo}
    D_{pp}  \sim  3\frac{\delta v_{\rm turb}^2}{cL}\eta_{B}^{-1/2}\psi^{-3}p^2.
\end{equation}
\par

The spectral index map of A2255 in \cite{Botteon_A2255} is a mixture of flat ($\alpha_{\rm syn}\sim1.0$) and steep ($\alpha_{\rm syn}\sim2.0$) regions.
The flattest index is coincident with the arc-shaped structure, suggesting the particle energization by shocks.
On the other hand, the emission with steep radio indices comes from the diffuse envelope permeating the cluster volume \citep[see Fig.~S2 of][]{Botteon_A2255} .
The mean value of the index is around $\alpha_{\rm syn}\approx1.6$. 
If we assume a scenario where particles emit in a homogeneous field, those steep indices suggest that the observing frequency is close to the steepening frequency, at which the spectrum starts to decline rapidly due to the cooled spectrum of CRe.

\par

In the turbulent reacceleration model of CRe, one can define the break energy $\gamma_{\rm b}$ as the energy where the cooling timescale $t_{\rm cool}$ becomes comparable to the acceleration timescale $t_{\rm acc}$. 
As shown in \citet{Cassano_2010A&A...509A..68C}, the steepening frequency $\nu_{\rm s}$ in the synchrotron spectrum appears at a factor $\xi\sim 5-7$ times larger than the break frequency $\nu_{\rm b}$ corresponding to $\gamma_{\rm b}$, i.e., $\nu_{\rm s}=\xi\nu_{\rm b}$.
The cooling time of ultra-relativistic CRe due to synchrotron and inverse-Compton radiation can be written as $t_{\rm cool} = 6\pi m_{e}c\gamma^{-1}/(\sigma_{\rm T}(B^2+B_{\rm CMB}^2))$, where $\gamma$ is the Lorenz factor of the particle, $\sigma_{\rm T}$ is the Thomson cross section and $B_{\rm CMB} = 3.25(1+z)^2\mu{\rm G}$ is the inverse-Compton (IC) equivalent field.
Note that 100 MHz emission is mainly produced by the synchrotron radiation of CRe with $p = \sqrt{\gamma^2-1}\sim10^4$, and the Coulomb loss is negligible for those CRe.
Using the magnetic field estimated from the dynamo model of Eq.~(\ref{eq:dynamo}), the cooling timescale at $\nu_{\rm b}$ can be estimated as
\begin{eqnarray}\label{eq:t_cool}
    t_{\rm cool}  &=&  \frac{\sqrt{27\pi em_ec}}{\sigma_{\rm T}}\frac{B^{1/2}}{B^2+B_{\rm CMB}^2}\xi^{\frac{1}{2}}\nu_{\rm s}^{-\frac{1}{2}}, \nonumber \\
    & \sim &  450~{\rm Myr} \left(\frac{\eta_B}{0.05}\right)^{\frac{1}{4}}(1+z)^{-\frac{5}{2}}\left(\frac{\mu}{0.59}\right)^{\frac{1}{4}}\left(\frac{n_{\rm ICM}}{10^{-4}~{\rm cm}^{-3}}\right)^{\frac{1}{4}} \nonumber \\
    &\;&\; \times \left(\frac{\delta v_{\rm turb}}{300~{\rm km/s}}\right)^\frac{1}{2}\left(\frac{\xi}{7}\right)^{\frac{1}{2}} \left(\frac{\nu_{\rm o}}{100~{\rm MHz}}\right)^{-\frac{1}{2}},
\end{eqnarray}
where $\mu$ is mean molecular weight, $\nu_{\rm o} = \nu_{\rm s}/(1+z)$ is the observing frequency.
The cooling is dominated by the radiation through the IC scattering ($B^2\ll B_{\rm CMB}^2$) in the cluster outskirts, so we neglected the $B^2$ term in the denominator.
The acceleration timescale should be comparable to this value to sustain the emission observed at the LOFAR frequency.
\par

\begin{figure}
	\includegraphics[width=\columnwidth]{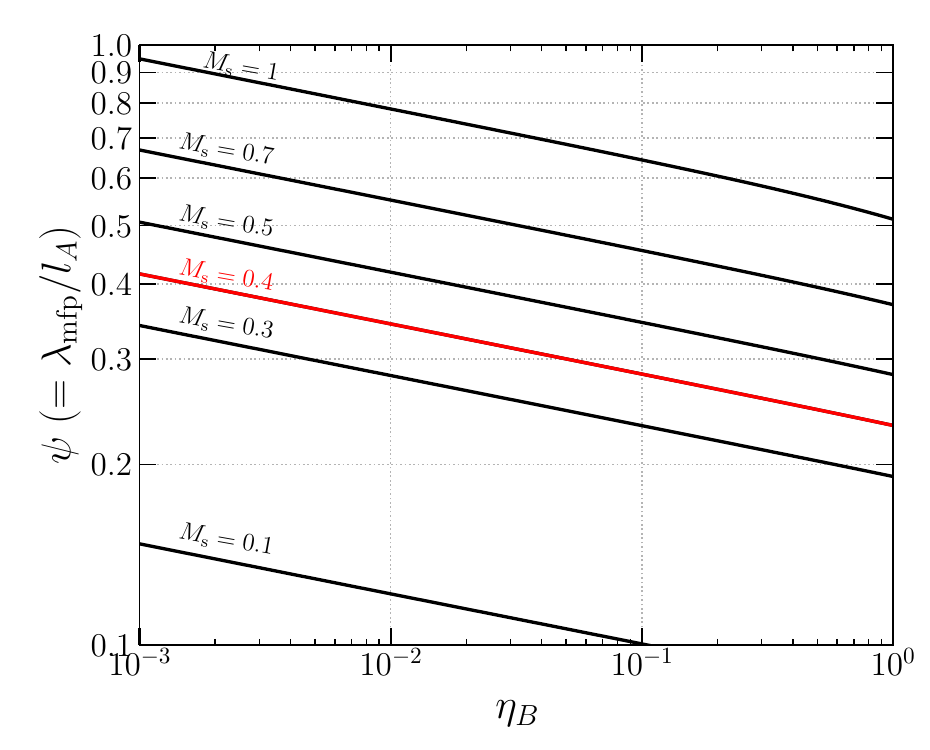}
    \caption{Relation between $\eta_B$ and $\psi$, required for the observing frequency of $\nu_{\rm o}=50$~MHz and $\xi=7$. Lines distinguish the results for the different turbulent velocities at $L=0.16~{\rm Mpc}$; $M_{\rm s} = 0.1,0.3,0.4,0.5,0.7$ and 1.0 (from bottom to top). For the ICM properties, we adopted typical values at the cluster periphery found in MHD simulations, $n_{\rm ICM}=3\times10^{-5}~{\rm cm}^{-3}$ and $T_{\rm ICM}=4$ keV. The source at $z=0$ is assumed.}
    \label{fig:etaB_psi}
\end{figure}

Comparing Eq.~(\ref{eq:t_cool}) with the acceleration timescale $t_{\rm acc} = p^2/(4D_{pp})$ obtained from Eq.~(\ref{eq:Dpp_dynamo}), one can relate two fundamental parameters, $\psi$ and $\eta_B$.
Fig.~\ref{fig:etaB_psi} shows the relation between $\eta_B$ and $\psi$ in the case of $\nu_{o} = 50~{\rm MHz}$ and $z=0$. 
We adopted the typical value for the ICM density and temperature in cluster outskirts; $n_{\rm ICM}=3\times10^{-5}~{\rm cm}^{-3}$ and $T_{\rm ICM}=4$ keV.
Adopting $\eta_B\sim0.05$ and the typical turbulent flux found in MHD simulation, $F_{\rm turb}\approx10^{44}~{\rm erg/s/Mpc^3}$ (Fig.~\ref{fig:F_turb}), we find that the observed steep spectrum at LOFAR frequencies can be explained with the mfp of CRe $\psi\sim0.5$ of the Alfv\'en scale.
This value is consistent with that adopted by \cite{BL16} and \citet{BV20} to explain radio halos and bridges.
The mfp comparable to $l_{\rm A}$ in turbulence reconnection is supported by the studies of tracers in MHD simulations \citep[][]{Kowal_2011ApJ...735..102K,Kowal_2012PhRvL}.\footnote{
more recent theoretical attempts have also investigated the role of mirroring on particle diffusion and acceleration \citep[][]{Lazarian_Xu_2021,Lazarian_Xu_2023}.
}
\par

\begin{figure}
	\includegraphics[width=\columnwidth]{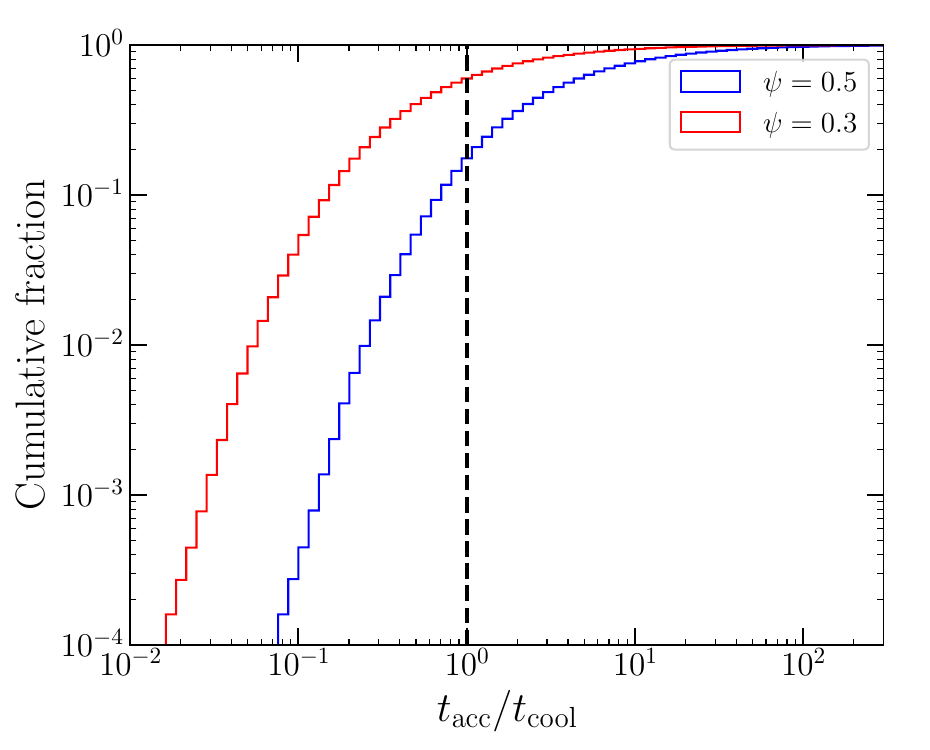}
    \caption{Cumulative number count of cells as a function of $t_{\rm acc}/t_{\rm cool}$ in the case of $\psi = 0.3$ (red) and 0.5 (blue).
    We adopted the same parameters as Fig.~\ref{fig:etaB_psi} ($\nu_{\rm o}$ = 50 MHz, $\xi = 7$, and $\eta_B = 0.05$).
    The vertical dotted line shows $t_{\rm acc}=t_{\rm cool}$.
    }
    \label{fig:tacc_tcool}
\end{figure}

One can calculate the number of cells that are important for the radio emission by comparing $t_{\rm acc}$ and $t_{\rm cool}$ measured in the extracted cubic volume.
Using Eqs.~(\ref{eq:Dpp_dynamo}) and (\ref{eq:t_cool}), the scaling relation for the fraction of those timescales can be expressed as
\begin{equation}\label{eq:tacc_vs_tcool}
 \frac{t_{\rm acc}}{t_{\rm cool}} \propto \rho^{-1/4}\delta v_{\rm turb}^{-5/2}L\xi^{-1/2}\psi^{3}\eta_{B}^{1/4},
\end{equation}
In Fig.~\ref{fig:tacc_tcool}, we show the cumulative fraction of $t_{\rm acc}/t_{\rm cool}$.
When $t_{\rm acc}<t_{\rm cool}$, the acceleration mechanism of Eq.(\ref{eq:Dpp}) can efficiently re-accelerate CRe that radiate synchrotron emission in the LOFAR frequencies.

The fraction of cells is $\sim$ 15\% adopting a reference value $\psi = 0.5$, while it increases to $\gtrsim$50 \% for $\psi<0.2-0.3$.
The dependence on $\eta_{B}$ is rather weak (Eq.~(\ref{eq:tacc_vs_tcool})).
The fraction ranges form 25\% to 10\% for $\eta_{B} = 0.01-0.1$, for a fixed $\psi = 0.5$.
Such a significant fraction implies a volume-filling emission.
One also needs to take care about the duration of reacceleration, since the reacceleration is important only when it lasts for several times longer than $t_{\rm acc}$.
The filling factor of the radio-emitting cells will be revisited in Sect.~\ref{sec:FP}.
\par


\section{Fokker--Planck simulation}\label{sec:FP}
Next, we study the synchrotron spectrum in our dynamo reacceleration model with a numerical calculation.
We consider the situation that the pre-existing population of seed CRe is re-accelerated by incompressible turbulence and produces the observed radio emission by the synchrotron radiation.
As in Sect.~\ref{sec:simulation}, we consider the volume of $1.6^3~{\rm Mpc}^3$ located $1.2~{\rm Mpc}$ from the center, and the distribution of the seed CRe is proportional to the ICM number density.
We solve the Fokker-Planck equation of CRe of the following form \citep[e.g.,][]{Cassano_Brunetti_2005}:
\begin{equation}\label{eq:FP}
\frac{\partial N_e(p,t)}{\partial t} = \frac{\partial}{\partial p}[\dot p N_e]+\frac{\partial}{\partial p}\left[D_{pp}\frac{\partial N_e}{\partial p}-\frac{2}{p}N_eD_{pp}\right]+Q_e(p,t),
\end{equation}
where $N_e(p,t)$ is the spectrum of CRe at time $t$, $\dot p$ represents the momentum loss per unit time, $D_{pp}$ is the momentum diffusion coefficient, and $Q_e$ is the injection spectrum of CRe.
The momentum loss rate $\dot p$ includes the effect of radiative(synchrotron and IC) and Coulomb losses, and $D_{pp}$ is calculated with Eq.~(\ref{eq:Dpp}).
We neglect the injection of the secondary electrons from inelastic $pp$ collisions of cosmic-ray protons.
Due to the low ICM density in the cluster outskirts, the contribution of the secondary electrons is expected to be smaller than that in classical radio halos (see Appendix~\ref{app:secondary} for further discussion).
\par

We neglect the term related to the spatial transport of CRs in Eq.~(\ref{eq:FP}), since it cannot be properly followed with a snapshot of the MHD simulation. 
However, we are considering the case where the mfp of CRe is comparable to the Alfv\'en scale ($l_{\rm A}\sim 1~{\rm kpc}$), and this leads to a large value of the spatial diffusion coefficient $D\sim 10^{31} ~{\rm cm^2}/{\rm s}$, as shown in Eq.~(\ref{eq:D_diff}).
The detailed investigation of the effect of spatial transport in the reacceleration model is left for future works using a Lagrangian tracer method.
In this exploratory work, we adopt following two approaches: a one-zone approximation (Sect.~\ref{subsec:onezone}), where we adopt the average value of the physical quantities found in the snapshot, and the calculation considering the variation of $F_{\rm turb}$ in each cell and the integration along the line of sight (LOS) (Sect.~\ref{subsec:cell}).




\subsection{One zone calculation}\label{subsec:onezone}
\begin{figure*}
    \plottwo{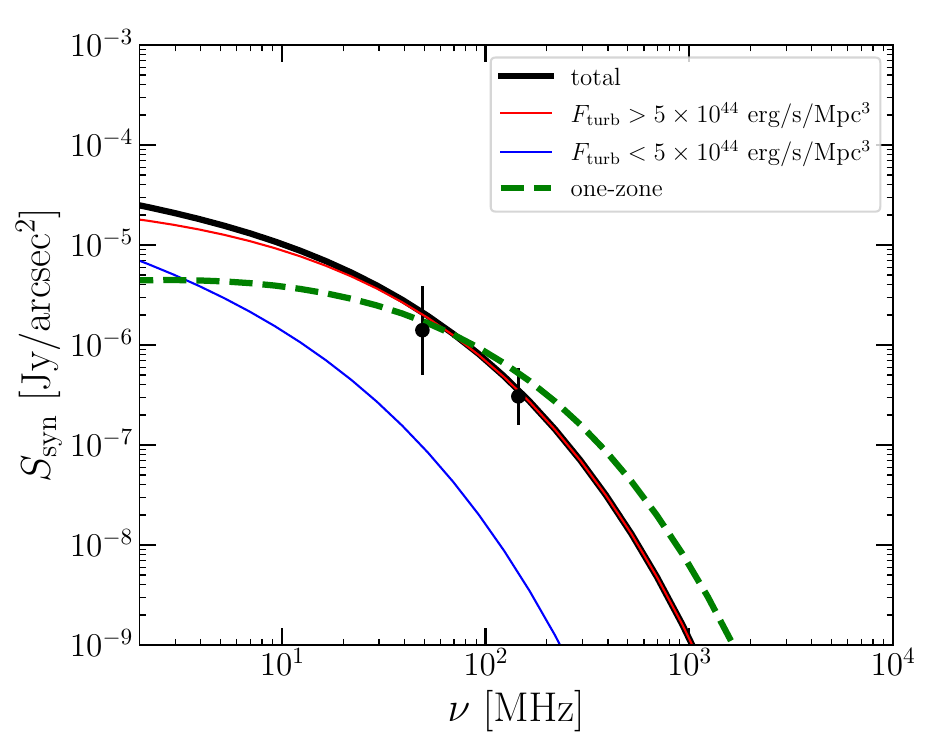}{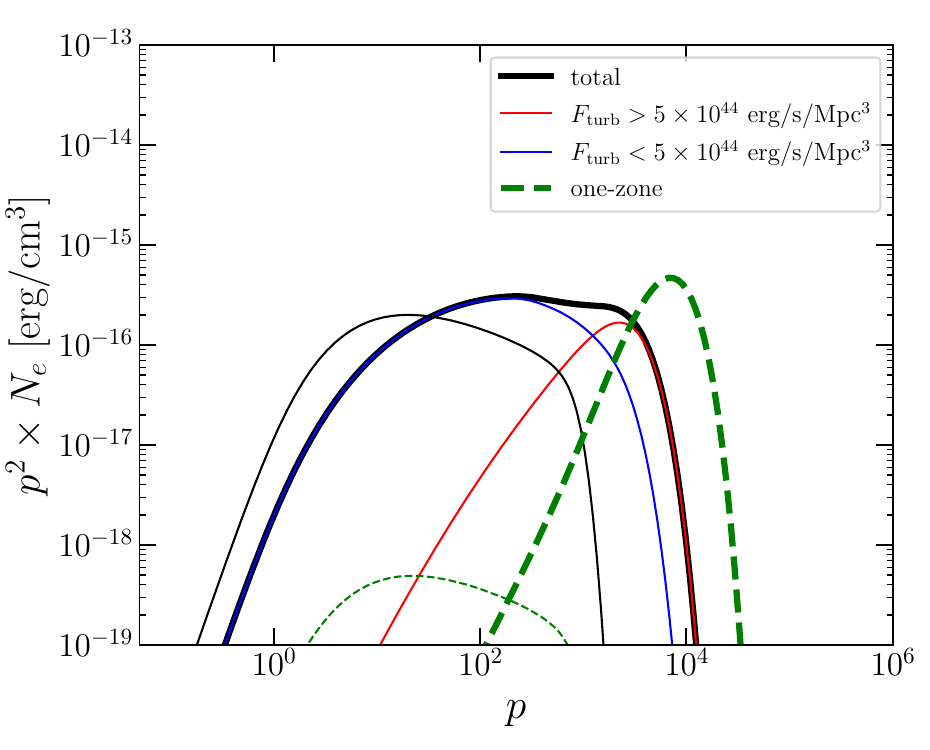}
    \caption{Left: The thick solid line shows the synchrotron spectrum in a single beam, typically seen in our $10^6$ cell calculation. For comparison, the spectrum in the one-zone model is shown with the dashed line. The data points show the typical value of the brightness measured at 1.5-2 Mpc from the cluster center \citep[][]{Botteon_A2255}. The error bars show the ranges of the brightness in this region including 1$\sigma$ error. We adopt $\psi = 0.3$ for the one-zone model, while $\psi = 0.6$ and for the cell-wise calculations. In both models, we assume $T_{\rm dur} = 3$ Gyr.
    The beam spectrum is decomposed into the contributions from the cells with $F_{\rm turb}<5\times10^{44}~{\rm erg/s/Mpc^3}$ (blue) and $F_{\rm turb}>5\times10^{44}~{\rm erg/s/Mpc^3}$ (red). 
    Right: CRe spectra in the same calculations.
    Spectra before and after the reacceleration is shown with thin and thick black lines, respectively. The dashed lines show the spectra in the one-zone model.}
    \label{fig:spectrum}
\end{figure*}

We first discuss the energy density of CRe required to reproduce the observed emission under the one-zone approximation.
We adopt the mean values in the simulation box for the physical quantities; $n_{\rm ICM} = 6\times10^{-5}~{\rm cm^{-3}}$, $T_{\rm ICM}=4.5~$keV, $v_{\rm sol}=320~{\rm km/s}$ ($M_{\rm s}\approx0.4$), and $t_{\rm eddy} = 0.53~{\rm Gyr}$.
We adopt $\eta_B = 0.05$, which corresponds to $B = 0.24\mu{\rm G}$ (Eq.~(\ref{eq:dynamo})). We summarize the values of parameters in Tab.~\ref{tab:params}.
As an initial condition, a cooled spectrum of seed CRe is calculated by integrating Eq.(\ref{eq:FP}) for 2 Gyr with $D_{pp}=0$ and $Q_e(p,t)\propto p^{-\alpha_{\rm inj}}\delta(t)$, where $\alpha_{\rm inj}=2.2$ and $\delta(t)$ is the Dirac delta function.
Due to the radiative and the Coulomb cooling, the seed CRe has steep cut-offs around $p_{\rm min}\sim10$ and $p_{\rm max}\sim10^3$ (Fig.~\ref{fig:spectrum} right panel).
The normalization of the initial spectrum is treated as a free parameter.
\par

\begin{table}\label{tab:params}
    \centering
    \caption{Parameters adopted in the FP calculations}
    \begin{tabular}{c|cc}
    \hline
       & Sect.~\ref{subsec:onezone} & Sect.~\ref{subsec:cell} \\
    \hline
       $\psi$ & 0.3 & 0.6 \\
       $\eta_B$ & 0.05 & 0.05 \\
       $T_{\rm dur}$ [Gyr] & 3 & 3 \\
       $L$ [kpc] & 160 & 160 \\
       $F_{\rm turb}$ [erg/s/Mpc$^3$] & $1.1\times 10^{44}$& --\tablenotemark{a}  \\
       $n_{\rm ICM}$ [cm$^{-3}$] & $6\times10^{-5}$ & --\tablenotemark{a} \\
       $T_{\rm ICM}$ [keV] & 4.5 & --\tablenotemark{a} \\
       $B$ [$\mu$G] & 0.24\tablenotemark{b} & --\tablenotemark{a,b} \\
    \hline
    \end{tabular}
    \tablenotetext{a}{We use the values found each simulated cell.}
    \tablenotetext{b}{The magnetic field is calculated with Eq.~(\ref{eq:dynamo}).}
\end{table}

As seen from Fig.~\ref{fig:etaB_psi}, for $M_{\rm s} \sim 0.4-0.7$, a value $\psi \sim 0.3-0.5$ is compatible with the observed emission.
This is confirmed by numerical calculation in Fig.~\ref{fig:spectrum} (left panel), where we plot the synchrotron spectrum in the one-zone model with a dashed line.
While the spectral shape is mainly determined by the parameters $\eta_{B}$ and $\psi$ (Eq.~(\ref{eq:tacc_vs_tcool})), the normalization (brightness) depends on the initial spectrum $N_e(p,0)$ and the duration of reacceleration $T_{\rm dur}$.
In this section, we assume that $T_{\rm dur}$ is sufficiently long and the CRe spectrum reaches a steady state due to the balance between the reacceleration and the radiative cooling (Eq.~(\ref{eq:FP})).
We find that this steady state is achieved at $T_{\rm dur} \geq 3~{\rm Gyr}$.
The duration of $T_{\rm dur} = 3~{\rm Gyr}$ is comparable to the dynamical timescale of the ICM in the simulated region, so the mean value of the turbulent flux $F_{\rm turb}$ can evolve in this timescale. Thus, the assumption of constant $F_{\rm turb}$ during the FP calculation should be considered as a rough approximation.
\par

We determine the normalization of the initial spectrum as the synchrotron brightness at 49MHz matches the observed value.
The data points in Fig.~\ref{fig:spectrum} show the range of the brightness of the diffuse envelope measured in 1.5-2 Mpc from the center of A2255 \citep[][]{Botteon_A2255}.

\par

We define the efficiency of the reacceleration, $\eta_{\rm acc}$, as the fraction of the turbulent kinetic flux that turns into the increase of the CRe energy density and the radiation:
\begin{equation}\label{eq:eta_acc}
    \eta_{\rm acc}F_{\rm turb} \equiv \frac{d\epsilon_{\rm CRe}}{dt} + \varepsilon_{\rm rad},
\end{equation}
where $\epsilon_{\rm CRe}$ is the CRe energy density and $\varepsilon_{\rm rad} = \varepsilon_{\rm syn} + \varepsilon_{\rm IC}$ is the sum of the frequency-integrated emissivities of synchrotron and IC radiations. 
In the steady state of $N_e$, the first term on the right-hand side is negligible.
The synchrotron emissivity $\varepsilon_{\rm syn}$ is calculated in the range of $10^4-10^{10}$ Hz, and $\varepsilon_{\rm IC}$ is calculated with $\varepsilon_{\rm IC} = (B_{\rm cmb}/B)^2\times\varepsilon_{\rm syn}$.
Since $B\ll B_{\rm cmb}$, $\varepsilon_{\rm rad}$ is dominated by $\varepsilon_{\rm IC}$. 
\par


In Fig.~\ref{fig:spectrum} (dashed line), we show the CRe energy spectrum in the steady state.
The energy is dominated by 100 MHz-emitting particles and the particles with $p<10^3$ are not energetically important.
We find $\epsilon_{\rm rad}\approx 7 \times 10^{-30}$ erg/s/cm$^3$ and $\eta_{\rm acc} \approx 2\%$, which is consistent with the estimate in \citet{Botteon_A2255}.
Due to the assumption of stationary conditions, this result is independent of the initial spectrum.

\par

We find that the spectrum in Fig.~\ref{fig:spectrum} is reasonably reproduced by considering a range of values of $\psi$ around the reference value $\psi\sim0.5$.
On the other hand, the assumption of a mfp that is significantly reduced, $\psi<0.2$, generates spectra that are too hard compared to the observation. 
Note that this result is based on the assumption that the turbulence and physical parameters in the simulated ICM are representative of the external regions in A2255.


\subsection{projection along the LOS}
\label{subsec:cell}
Next, we consider the distribution of turbulent energy in the box and study how the emission from highly turbulent cells affects the spectrum integrated along the LOS.
To consider the CRe spectrum in each simulated cell, we calculate the FP equation for various $F_{\rm turb}$ found in the simulated box.
We make a histogram of $F_{\rm turb}$ with 160 bins equally spaced in the logarithmic scale in the range of $10^{38}~{\rm erg/s/Mpc^3}< F_{\rm turb} < 10^{46}~{\rm erg/s/Mpc^3}$ (Fig.~\ref{fig:F_turb}). 
For simplicity, we adopt the same values of dynamo $B$ (Eq.~(\ref{eq:dynamo})) and $D_{pp}$ (Eq.~(\ref{eq:Dpp_dynamo})) for the cells in the same $F_{\rm turb}$ by calculating the mean values of $\rho_{\rm ICM}$ and $v_{\rm turb}$ in each $F_{\rm turb}$ bin.
The synchrotron brightness is calculated by integrating the emissivity of 100 cells (corresponds to 1.6~Mpc) along the axis of the simulation.
In one projection, there are 100$\times$100 LOS.
\par

We assume the same initial spectrum as in Sect.~\ref{subsec:onezone}, i.e., the spectrum with cut-offs at $p_{\rm min}\sim10$ and $p_{\rm max}\sim10^3$.
As in the previous section, we consider a sufficiently long $T_{\rm dur} = 3$ Gyr to ensure that the steady state due to the balance between cooling and reacceleration is achieved in many of the cells.
Note that the turbulent flux in each cell would change in a few eddy turnover time, $t_{\rm eddy}\sim0.3-1~{\rm Gyr}$.
When $T_{\rm dur}$ is short and the steady state is not achieved, the result may depend on the initial condition.
This point is further discussed in Appendix~\ref{app:initial}, where we show that the observed emission can be explained with a different combination of $T_{\rm dur}$ and the initial condition.
\par



The non-dimensional parameters in our model, $\psi$ (Eq.~(\ref{eq:Dpp})) and $\eta_B$ (Eq.~(\ref{eq:dynamo})), are assumed to be constant over cells.
As a test case, we adopt the same $\eta_{B} = 0.05$ as the one zone calculation.
We use a snapshot of the simulation and do not consider the evolution of the background fluid.
We assume that the initial energy density of the CRe in each cell is proportional to the thermal energy density i.e., $\epsilon_{\rm CRe}(t=0) \propto \epsilon_{\rm ICM}$.
We neglect the particle transport between cells during the calculation.
Those simplification reduces the computational cost and enables us to calculate the spectrum in every cell in the box for several Gyrs.
The study on the impact of CR transport is left to future works featuring the Lagrangian tracer approach (Sect.~\ref{sec:limitation}).
\par


We calculate the distribution of the spectral index and compare it with the observation by \citet{Botteon_A2255}.
The spectral index and its statistical properties would depend on the beam size and sensitivity of the observation, so we consider those of LOFAR.
The beam size in \cite{Botteon_A2255} is 35", which corresponds to almost 3 times the cell size at the redshift of A2255, so we calculate the spectrum of one beam by summing up the intensities of $3\times3$ neighboring LOS.
Each simulated beam consists of the emission from 900 cells.
Following \citet{Botteon_A2255}, we introduce a cut in our simulated data, neglecting the beams that has 145 MHz brightness below $2\sigma_{\rm HBA}$, where $\sigma_{\rm HBA}=200\mu{\rm Jy}$ is the r.m.s. noise per beam of the LOFAR HBA band.
\par



\begin{figure}
	\includegraphics[width=\columnwidth]{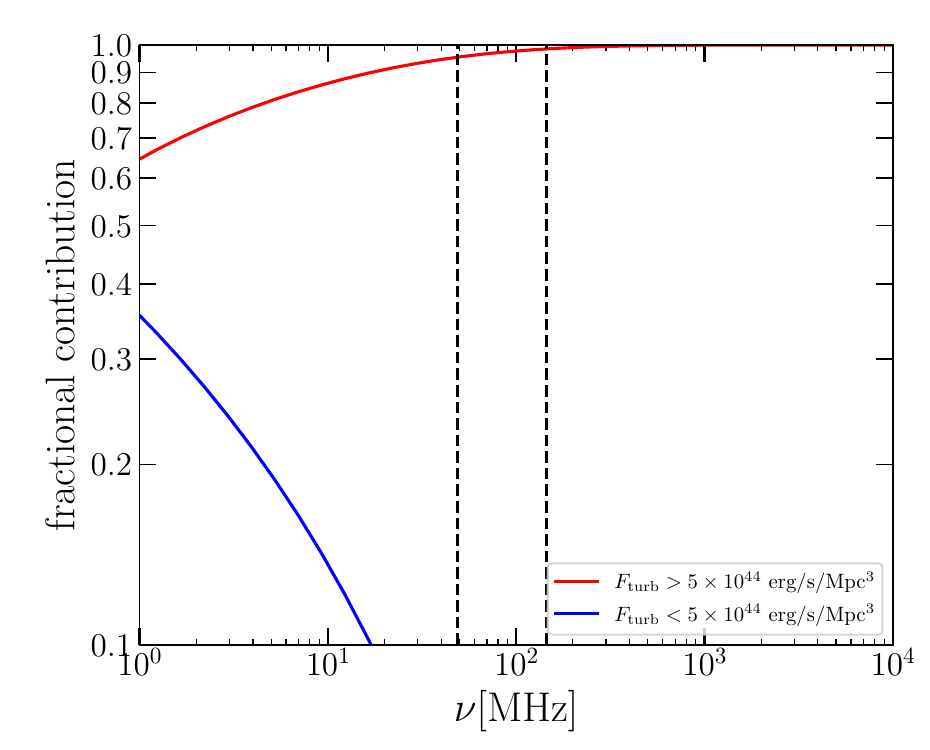}
    \caption{Fractional contribution from $F_{\rm turb}>5\times10^{44}~{\rm erg/s/Mpc^3}$ (orange) and $F_{\rm turb}<5\times10^{44}~{\rm erg/s/Mpc^3}$ (blue) cells to the flux shown in Fig.~\ref{fig:spectrum}. The vertical dashed lines show the LBA and HBA frequencies of LOFAR.}
\label{fig:frac_freq}
\end{figure}

In Fig.~\ref{fig:spectrum}, we show the typical synchrotron spectrum of beam detectable with the LOFAR sensitivity (black); here we specifically use $\psi = 0.6$.
Since more turbulent cells contribute more than less turbulent cells the results are not exactly consistent with those in the one-zone model (Sect.~\ref{subsec:onezone}) or with Fig~\ref{fig:etaB_psi}.
In fact, the model considering the LOS integration of  $\sim 10^3$ cells (in each beam) requires a slightly larger value of $\psi$ and consequently a slightly less efficient acceleration mechanism.

We decompose the spectrum into the contributions from $F_{\rm turb}>5\times10^{44}~{\rm erg/s/Mpc^3}$ (orange) and $F_{\rm turb}<5\times10^{44}~{\rm erg/s/Mpc^3}$ (blue) cells.
We confirm the trend seen in \citet{BV20};
at higher frequencies ($\nu>100$MHz), the highly turbulent cells dominate the emission, which occupies only a small fraction (4.5\%) of the volume, while the emission is more volume-filling at lower frequencies.
In Fig.~\ref{fig:frac_freq}, we show the contribution from the turbulent cells as a function of frequency.
CRe spectra are compared in Fig.~\ref{fig:spectrum}.
Unlike one-zone calculation, the overall spectrum has a bump around $p\sim100$.
CRe in the cells with smaller turbulent energy dominates the CRe energy, although they do not significantly contribute to the emission at $\sim100$ MHz (Fig.~\ref{fig:frac_freq}).
The typical momentum of CRe that corresponds to 100 MHz emission shifts to $p\sim2-5\times10^3$, since the magnetic field in $F_{\rm turb}\approx 5\times10^{44}$ erg/s/Mpc$^3$ cells is a factor $\sim$2 stronger than $B=0.24~\mu$G adopted in the one-zone model.
As discussed in Appendix~\ref{app:initial}, the contribution from low $F_{\rm turb}$ can be larger and the emission becomes more volume-filling under different assumption on the initial condition.
\par



In reality, the CR distribution would be smoothed by diffusion and/or streaming, and CRs would experience multiple reacceleration within the dynamical time.
In such a situation, the gap between cells with large $F_{\rm turb}$ and small $F_{\rm turb}$ would be reduced and the emission becomes more volume-filling.
This point would be further studied in future studies with a Lagrangian tracer method (see also Sect.~\ref{sec:limitation}).
\par

We find that 375 beams out of 1089 satisfy the criterion of detection, i.e., $S_{\rm HBA} > 2\sigma_{\rm HBA}$, so almost 34\% of the area of emission region can be covered by the LOFAR sensitivity.
This fraction increases with the efficiency of reacceleration (smaller $\psi$), as the contribution by the less turbulent cells ($F_{\rm turb}<5\times10^{44}~{\rm erg/s/Mpc^3}$) becomes more significant. 
However, we find that, for $\psi\leq0.3$, the typical spectrum starts to become flatter than the observed one.
\par


The value of $\eta_{\rm acc}$ (Eq.~(\ref{eq:eta_acc})) differs in cell to cell, and it becomes $\eta_{\rm acc}\approx0$ when $F_{\rm turb}$ is very small and the effect of the reacceleration is negligible.
To obtain a typical value of $\eta_{\rm acc}$ in this model, we calculate
\begin{equation} \label{eq:eta_acc_beam}
    \eta_{\rm acc} = \frac{\sum(\frac{d\epsilon_{\rm CRe}}{dt}+\varepsilon_{\rm rad})}{\sum F_{\rm turb}},
\end{equation}
where the sum is taken for the cells in each beam ($\sim10^3$ cells). We calculate the mean value of $\eta_{\rm acc}$ for the 375 beams and find $\langle\eta_{\rm acc}\rangle\approx 1\%$. 
Although there is a plenty of turbulent energy in $F_{\rm turb}>5\times10^{44}~{\rm erg/s/Mpc^3}$ cells, the occurrence of such cells is small (4.5\%).
The combination of those two result in $\eta_{\rm acc}\approx1\%$, which is slightly smaller than that found in the one zone model (Sect.~\ref{subsec:onezone}).
\par


\begin{figure}
	\includegraphics[width=\columnwidth]{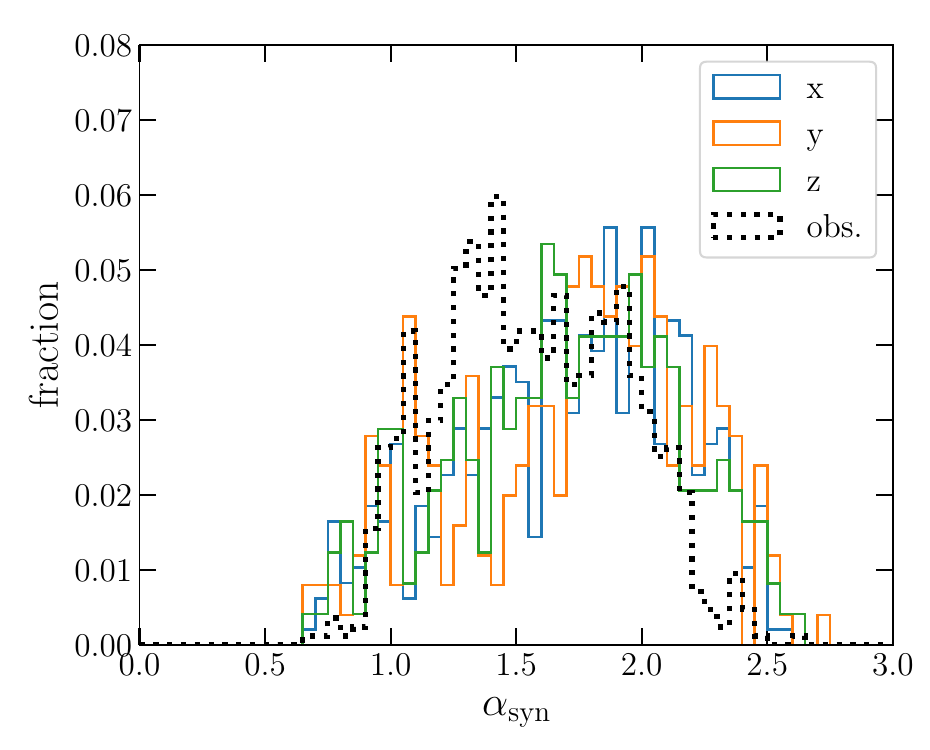}
    \caption{Distribution of spectral indices of the beams observable with the LOFAR sensitivity. 
    The vertical axis shows the fraction of beams with each $\alpha_{\rm syn}$. 
    The black dotted line shows the observation of \citet{Botteon_A2255}. 
    In our calculation, the total number of observable beams is $\approx 300$ (30\% of the area), while the observational data shows the distribution of  836 pixels. 
    For the model parameters, we adopted $\eta_B = 0.05$ and $\psi = 0.6$. 
    The reacceleration is calculated for $T_{\rm dur} = 3$ Gyr, and the beam size is assumed to be 35".
    We tested the projection along three different axes ($x,y,$ and $z$, distinguished by colors).}
    \label{fig:spix}
\end{figure}

Fig.~\ref{fig:spix} shows the distribution of spectral index in each beam calculated with $\psi = 0.6$ and $\eta_B = 0.05$. 
The black dashed line shows the result reported in \citet{Botteon_A2255}.
The mean value of the index in our calculation is $\langle\alpha\rangle\approx1.7$, and the difference in the results of three different projections is marginal.
Our results are in line with the observation of diffuse radio emission enveloping A2255.
\par

\section{limitations}\label{sec:limitation}

Clearly, the complex morphology observed by LOFAR suggests that at least shocks and turbulence contribute to the emission. 
A full modeling of A2255 is however beyond the aim of the present work.
We have explored the possibility of turbulent reacceleration in the context of the specific model \citet{BL16}. 
We find that the steep-spectrum diffuse emission observed at 1-2 Mpc distance from the center of A2255 can be explained using parameters ($\eta_B$ and $\psi$) that are in line with previous literature \citep[][]{BL16,BV20}.
In this case, the spatial diffusion coefficient of the CRe becomes $D \sim 10^{31} {\rm cm}^2{\rm s}^{-1}$ (Eq.~(\ref{eq:D_diff})), i.e., CRe can diffuse over 450 kpc within 1 Gyr. 
Within the acceleration time of radio-emitting CRe $t_{\rm acc}\sim 500~{\rm Myr}$ (Eq.~(\ref{eq:t_cool})), the diffusion length is $\sim 300$ kpc.
In addition, CRe can be spread and mixed by turbulent motion.
This implies that CRe can travel $\sim$30 cells during reacceleration, making the distribution smoother.
Since the cells that efficiently accelerate CRs appear in every $\sim 3^3$ cell (i.e., $\sim 4\%$), the diffusion is fast enough to fill the space between turbulent cells.
\par

We tested two cases: the one-zone model in Sect.~\ref{subsec:onezone} and the cell-wise calculation in Sect.~\ref{subsec:cell}.
In the former case, we adopt the average values of the physical quantities found in the simulated box to calculate the FP equation.
In the latter case, we neglected the transport within a short duration of the reacceleration, and calculated the reacceleration using the local value of the turbulent flux in each cell. 
A future study with a Lagrangian tracer method will be important to discuss the evolution of the CR spectrum and spatial distribution due to the combination of the reacceleration, the spatial diffusion, and advection. 
An observation with higher angular resolution would also be important to study the correlation between the flat spectrum and large turbulent kinetic flux predicted in Sect.~\ref{subsec:cell}, or the gradient of the spectral index around the turbulent region due to the diffusion.
\par

Concerning the initial condition, we have assumed that there is a cooling phase with $D_{pp}=0$ before the reacceleration starts, as in many models of the giant radio halo \citep[e.g.,][]{Brunetti_2001MNRAS.320..365B,Nishiwaki_2022ApJ}.
However, CRs can be gently re-accelerated for several Gyrs by the modest level of turbulence continuously driven by mass accretion.
This would imply that there is no clear onset of the mega halo emission, unlike the classical halos which are supposed to be driven by the mergers of clusters \citep[e.g.,][]{Cassano2016}.
In Appendix~\ref{app:initial}, we assume a different initial condition, assuming that the reacceleration was working before the epoch of the snapshot.
In this case, the observed emission can be explained by a an efficiency that is slightly reduced, $\eta_{\rm acc} \approx 1.7\%$.

\par

The tracer approach is also important for that issue. 
Following the energy gain and loss of CRe with a tracer method in a simulated galaxy cluster, \citet{Beduzzi_2023arXiv230603764B} demonstrated that $\approx 22 - 57\%$ of the mega halo region ($0.4R_{500}<r<R_{500}$) is filled with radio-emitting CRe. 
Although the details of the simulation setup and the definition of the volume filling factor are different from our study, their result suggests that the cluster-scale diffuse emission is produced through the multiple episodes of turbulent reacceleration.
\par


When extracting the peripheral region of the simulated cluster in Sect.~\ref{sec:simulation}, we choose the particular sector where a large-scale accretion can be seen (Fig.~\ref{fig:sim}), motivated by the fact that the most turbulent region dominates the emission in our model (Sect.~\ref{sec:FP}).
We note that the typical turbulent energy can be smaller in other sectors without such an accretion feature.
Considering that the properties of the gas and turbulence seen in the MHD simulation can be different from those in A2255, the parameters reported in this study are basically indicative values.

\par


\section{Conclusions}\label{sec:concl}
Recent LOFAR observations reported the presence of diffuse radio emission permeating the volume of galaxy clusters up to the virial radius.
This emission is termed a mega halo, and its mechanism is still unclear.
\par

The diffuse radio emission in A2255 has a very large extension, enveloping the example of classical halo and relics reported in previous observations \citep[e.g.,][]{Botteon_A2255}.
The complex radio morphology is likely generated through a mix of different processes, e.g., particle acceleration by shocks and turbulence. 
The diffuse emission permeating the cluster volume on large scales is a candidate for turbulent reacceleration. In this paper, we have explored this hypothesis.
\par


Recent MHD simulations observe that the turbulent energy is dominated by the solenoidal component in the cluster outskirts, so we focus on the acceleration mechanism presented by \citet{BL16}, where particles are accelerated by the interaction between particles and magnetic field lines diffusing in a super-Alfv\'enic solenoidal turbulent flow.
The acceleration efficiency $D_{pp}$ depends on the mfp of the CR particle $\psi \equiv \lambda_{\rm mfp}/l_{\rm A}$, which is treated as a free parameter (Eq.~(\ref{eq:Dpp})).
\par


We use a snapshot of a simulated cluster mimicking A2255 and study the property of turbulence and magnetic field in the cluster outskirts.
We extract a cubic volume of a region in the cluster periphery, where the prominent feature of mass accretion can be seen (Fig.~\ref{fig:sim}).
Since the spatial resolution of the simulation is not sufficient to resolve the small-scale dynamo at the Alfv\'en scale, the magnetic field is estimated in a post-process.
We assume that $\eta_B \approx 0.05$ of the turbulent flux is consumed as the dynamo and obtain $B\approx0.2~\mu{\rm G}$, which is roughly two times larger than the field found in the simulation.
Comparing the efficiency of reacceleration and radiative cooling, we find that the diffuse emission in A2255 can be explained with a fiducial value of $\psi$($\sim 0.5$) that was derived from considerations based on mirroring of electrons in a super-Alfv\'enic flow \citep[][]{BL16,BV20}.
\par

The emission spectrum is calculated by numerically integrating the FP equation (Eq.~\ref{eq:FP}) with the quantities found in the MHD simulation.
Before the reacceleration, CRe is accumulated around $p\sim10^2$ due to the radiative and Coulomb cooling before the reacceleration.
In Sect.~\ref{subsec:onezone}, we adopted the one-zone approximation and used the average quantities in the simulation box.
We find that the reacceleration by the solenoidal turbulence is efficient enough to produce a CRe spectrum peaked at $p\sim10^4$, corresponding to the synchrotron frequency of $\sim$100 MHz in the $\sim 0.1\mu{\rm G}$ field.
The acceleration efficiency is $\eta_{\rm acc}\approx 0.02$, in line with the estimate in \citet{Botteon_A2255}.
\par

We also calculate the FP equation in $10^6$ cells in the simulation box, assuming that the seed CRe is uniformly distributed.
We find that 100 MHz emission is dominated by the cells with large turbulent kinetic flux ($F_{\rm turb}>5\times10^{44}~{\rm erg/s/Mpc^3}$), which fill only a few \% of the volume.
Considering the LOS integration within the beam size of LOFAR, we find that a large fraction of the beams ($\gtrsim30\%$) can be detected with the LOFAR sensitivity.
In this model, only 1\% of the turbulent energy needs to be consumed for the particle acceleration in the turbulent cells.
Our model predicts that the emission will be more volume-filling when observed with higher sensitivity at lower frequencies.

The reported parameters and the derived efficiencies are indicative values, as the simulations do not necessarily reproduce the turbulent properties of A2255. For this reason, our study simply suggests that there is room for turbulent reacceleration to contribute to the observed emission.

\par



In reality, CRe in each cell would be mixed by the streaming and/or diffusion, although we neglect those effects in the current study.
As discussed in Sect.~\ref{sec:limitation}, the mfp comparable to the Alfv\'en scale leads to a strong diffusion with $D\sim 10^{31}~{\rm cm^2/s}$, and CRe can diffuse over a few 100 kpc within $\sim$1 Gyr.
A method that can incorporate those effects, such as the application of a Lagrangian tracer method \citep[e.g.][]{2023A&A...669A..50V}, will be important for more accurate modeling.
\par


\section*{Acknowledgements}
The authors thank the anonymous referee for the useful comments.
We acknowledge Andrea Botteon, Katsuaki Asano, and Takumi Ohmura for the fruitful discussions.
This research was supported by FoPM, WINGS Program, the University of Tokyo.
K. N. acknowledges the support by JSPS KAKENHI grants No. JP23KJ0486.
Numerical computations were carried out on PLEIADI at IRA, INAF (http://www.pleiadi.inaf.it). F.\,V. acknowledges financial support from the Horizon 2020 program under the ERC Starting Grant \texttt{MAGCOW}, no. 714196, from the Cariplo "BREAKTHRU" funds (Rif: 2022-2088 CUP J33C22004310003), and the usage of computing time  through the John von Neumann Institute for Computing (NIC) on the GCS Supercomputer JUWELS at J\"ulich Supercomputing center (JSC), under project "radgalicm2".
In this work, we used the \texttt{enzo} code (\hyperlink{http://enzo-project.org}{http://enzo-project.org}), the product of a collaborative effort of scientists at many universities and national laboratories. 
The authors gratefully acknowledge the Gauss center for Supercomputing e.V. (www.gauss-center.eu) for supporting this project by providing computing time through the John von Neumann Institute for Computing (NIC) on the GCS Supercomputer JUWELS at J\"ulich Supercomputing center (JSC), under project "radgalicm2" (P.I. F. Vazza). 

\bibliography{MegaHalo}{}

\begin{thebibliography}{}
\expandafter\ifx\csname natexlab\endcsname\relax\def\natexlab#1{#1}\fi
\providecommand{\url}[1]{\href{#1}{#1}}
\providecommand{\dodoi}[1]{doi:~\href{http://doi.org/#1}{\nolinkurl{#1}}}
\providecommand{\doeprint}[1]{\href{http://ascl.net/#1}{\nolinkurl{http://ascl.net/#1}}}
\providecommand{\doarXiv}[1]{\href{https://arxiv.org/abs/#1}{\nolinkurl{https://arxiv.org/abs/#1}}}

\bibitem[{{Ackermann} {et~al.}(2014){Ackermann}, {Ajello}, {Albert},
  {Allafort}, {Atwood}, {Baldini}, {Ballet}, {Barbiellini}, {Bastieri},
  {Bechtol}, {Bellazzini}, {Bloom}, {Bonamente}, {Bottacini}, {Brandt},
  {Bregeon}, {Brigida}, {Bruel}, {Buehler}, {Buson}, {Caliandro}, {Cameron},
  {Caraveo}, {Cavazzuti}, {Chaves}, {Chiang}, {Chiaro}, {Ciprini}, {Claus},
  {Cohen-Tanugi}, {Conrad}, {D'Ammando}, {de Angelis}, {de Palma}, {Dermer},
  {Digel}, {Drell}, {Drlica-Wagner}, {Favuzzi}, {Franckowiak}, {Funk}, {Fusco},
  {Gargano}, {Gasparrini}, {Germani}, {Giglietto}, {Giordano}, {Giroletti},
  {Godfrey}, {Gomez-Vargas}, {Grenier}, {Guiriec}, {Gustafsson}, {Hadasch},
  {Hayashida}, {Hewitt}, {Hughes}, {Jeltema}, {J{\'o}hannesson}, {Johnson},
  {Kamae}, {Kataoka}, {Kn{\"o}dlseder}, {Kuss}, {Lande}, {Larsson},
  {Latronico}, {Llena Garde}, {Longo}, {Loparco}, {Lovellette}, {Lubrano},
  {Mayer}, {Mazziotta}, {McEnery}, {Michelson}, {Mitthumsiri}, {Mizuno},
  {Monzani}, {Morselli}, {Moskalenko}, {Murgia}, {Nemmen}, {Nuss}, {Ohsugi},
  {Orienti}, {Orlando}, {Ormes}, {Perkins}, {Pesce-Rollins}, {Piron}, {Pivato},
  {Rain{\`o}}, {Rando}, {Razzano}, {Razzaque}, {Reimer}, {Reimer}, {Ruan},
  {S{\'a}nchez-Conde}, {Schulz}, {Sgr{\`o}}, {Siskind}, {Spandre}, {Spinelli},
  {Storm}, {Strong}, {Suson}, {Takahashi}, {Thayer}, {Thayer}, {Thompson},
  {Tibaldo}, {Tinivella}, {Torres}, {Troja}, {Uchiyama}, {Usher},
  {Vandenbroucke}, {Vianello}, {Vitale}, {Winer}, {Wood}, {Zimmer}, {Fermi-LAT
  Collaboration}, {Pinzke}, \& {Pfrommer}}]{Ackermann_2014}
{Ackermann}, M., {Ajello}, M., {Albert}, A., {et~al.} 2014, \apj, 787, 18,
  \dodoi{10.1088/0004-637X/787/1/18}

\bibitem[{{Ackermann} {et~al.}(2016){Ackermann}, {Ajello}, {Albert}, {Atwood},
  {Baldini}, {Ballet}, {Barbiellini}, {Bastieri}, {Bechtol}, {Bellazzini},
  {Bissaldi}, {Blandford}, {Bloom}, {Bonino}, {Bottacini}, {Bregeon}, {Bruel},
  {Buehler}, {Caliandro}, {Cameron}, {Caragiulo}, {Caraveo}, {Casandjian},
  {Cavazzuti}, {Cecchi}, {Charles}, {Chekhtman}, {Chiaro}, {Ciprini},
  {Cohen-Tanugi}, {Conrad}, {Cutini}, {D'Ammando}, {de Angelis}, {de Palma},
  {Desiante}, {Digel}, {Di Venere}, {Drell}, {Favuzzi}, {Fegan}, {Fukazawa},
  {Funk}, {Fusco}, {Gargano}, {Gasparrini}, {Giglietto}, {Giordano},
  {Giroletti}, {Godfrey}, {Green}, {Grenier}, {Guiriec}, {Hays}, {Hewitt},
  {Horan}, {J{\'o}hannesson}, {Kuss}, {Larsson}, {Latronico}, {Li}, {Li},
  {Longo}, {Loparco}, {Lovellette}, {Lubrano}, {Madejski}, {Maldera},
  {Manfreda}, {Mayer}, {Mazziotta}, {Michelson}, {Mitthumsiri}, {Mizuno},
  {Monzani}, {Morselli}, {Moskalenko}, {Murgia}, {Nuss}, {Ohsugi}, {Orienti},
  {Orlando}, {Ormes}, {Paneque}, {Pesce-Rollins}, {Petrosian}, {Piron},
  {Pivato}, {Porter}, {Rain{\`o}}, {Rando}, {Razzano}, {Reimer}, {Reimer},
  {S{\'a}nchez-Conde}, {Sgr{\`o}}, {Siskind}, {Spada}, {Spandre}, {Spinelli},
  {Tajima}, {Takahashi}, {Thayer}, {Tibaldo}, {Torres}, {Tosti}, {Troja},
  {Vianello}, {Wood}, {Zimmer}, {Fermi-LAT Collaboration}, \&
  {Rephaeli}}]{Ackermann_2016}
---. 2016, \apj, 819, 149, \dodoi{10.3847/0004-637X/819/2/149}

\bibitem[{{Adam} {et~al.}(2021){Adam}, {Goksu}, {Brown}, {Rudnick}, \&
  {Ferrari}}]{Adam_2021}
{Adam}, R., {Goksu}, H., {Brown}, S., {Rudnick}, L., \& {Ferrari}, C. 2021,
  \aap, 648, A60, \dodoi{10.1051/0004-6361/202039660}

\bibitem[{{Akamatsu} {et~al.}(2017){Akamatsu}, {Mizuno}, {Ota}, {Zhang}, {van
  Weeren}, {Kawahara}, {Fukazawa}, {Kaastra}, {Kawaharada}, {Nakazawa},
  {Ohashi}, {R{\"o}ttgering}, {Takizawa}, {Vink}, \&
  {Zandanel}}]{Akamatsu_2017}
{Akamatsu}, H., {Mizuno}, M., {Ota}, N., {et~al.} 2017, \aap, 600, A100,
  \dodoi{10.1051/0004-6361/201628400}

\bibitem[{{Beduzzi} {et~al.}(2023){Beduzzi}, {Vazza}, {Brunetti}, {Cuciti},
  {Wittor}, \& {Corsini}}]{Beduzzi_2023arXiv230603764B}
{Beduzzi}, L., {Vazza}, F., {Brunetti}, G., {et~al.} 2023, arXiv e-prints,
  arXiv:2306.03764, \dodoi{10.48550/arXiv.2306.03764}

\bibitem[{{Beresnyak}(2012)}]{Beresnyak_2012}
{Beresnyak}, A. 2012, \prl, 108, 035002, \dodoi{10.1103/PhysRevLett.108.035002}

\bibitem[{{Beresnyak} \&
  {Miniati}(2016)}]{Beresnyak_Miniati_2016ApJ...817..127B}
{Beresnyak}, A., \& {Miniati}, F. 2016, \apj, 817, 127,
  \dodoi{10.3847/0004-637X/817/2/127}

\bibitem[{{Blasi} \& {Colafrancesco}(1999)}]{Blasi_1999APh....12..169B}
{Blasi}, P., \& {Colafrancesco}, S. 1999, Astroparticle Physics, 12, 169,
  \dodoi{10.1016/S0927-6505(99)00079-1}

\bibitem[{{Botteon} {et~al.}(2020{\natexlab{a}}){Botteon}, {van Weeren},
  {Brunetti}, {de Gasperin}, {Intema}, {Osinga}, {Di Gennaro}, {Shimwell},
  {Bonafede}, {Br{\"u}ggen}, {Cassano}, {Cuciti}, {Dallacasa}, {Gastaldello},
  {Mandal}, {Rossetti}, \& {R{\"o}ttgering}}]{Botteon_2020}
{Botteon}, A., {van Weeren}, R.~J., {Brunetti}, G., {et~al.}
  2020{\natexlab{a}}, \mnras, 499, L11, \dodoi{10.1093/mnrasl/slaa142}

\bibitem[{{Botteon} {et~al.}(2020{\natexlab{b}}){Botteon}, {Brunetti}, {van
  Weeren}, {Shimwell}, {Pizzo}, {Cassano}, {Iacobelli}, {Gastaldello},
  {B{\^\i}rzan}, {Bonafede}, {Br{\"u}ggen}, {Cuciti}, {Dallacasa}, {de
  Gasperin}, {Di Gennaro}, {Drabent}, {Hardcastle}, {Hoeft}, {Mandal},
  {R{\"o}ttgering}, \& {Simionescu}}]{Botteon_2020_A2255}
{Botteon}, A., {Brunetti}, G., {van Weeren}, R.~J., {et~al.}
  2020{\natexlab{b}}, \apj, 897, 93, \dodoi{10.3847/1538-4357/ab9a2f}

\bibitem[{{Botteon} {et~al.}(2022){Botteon}, {van Weeren}, {Brunetti}, {Vazza},
  {Shimwell}, {Br{\"u}ggen}, {R{\"o}ttgering}, {de Gasperin}, {Akamatsu},
  {Bonafede}, {Cassano}, {Cuciti}, {Dallacasa}, {Di Gennaro}, \&
  {Gastaldello}}]{Botteon_A2255}
{Botteon}, A., {van Weeren}, R.~J., {Brunetti}, G., {et~al.} 2022, Science
  Advances, 8, eabq7623, \dodoi{10.1126/sciadv.abq7623}

\bibitem[{Brunetti \& Jones(2014)}]{Brunetti_Jones_review}
Brunetti, G., \& Jones, T.~W. 2014, International Journal of Modern Physics D,
  23, 1, \dodoi{10.1142/S0218271814300079}

\bibitem[{{Brunetti} \& {Lazarian}(2007)}]{Brunetti_Lazarian_2007}
{Brunetti}, G., \& {Lazarian}, A. 2007, \mnras, 378, 245,
  \dodoi{10.1111/j.1365-2966.2007.11771.x}

\bibitem[{{Brunetti} \&
  {Lazarian}(2011{\natexlab{a}})}]{Brunetti_Lazarian_2011MNRAS.410..127B}
---. 2011{\natexlab{a}}, \mnras, 410, 127,
  \dodoi{10.1111/j.1365-2966.2010.17457.x}

\bibitem[{{Brunetti} \&
  {Lazarian}(2011{\natexlab{b}})}]{Brunetti_Lazarian_2011}
---. 2011{\natexlab{b}}, \mnras, 412, 817,
  \dodoi{10.1111/j.1365-2966.2010.17937.x}

\bibitem[{{Brunetti} \& {Lazarian}(2016)}]{BL16}
---. 2016, \mnras, 458, 2584, \dodoi{10.1093/mnras/stw496}

\bibitem[{{Brunetti} {et~al.}(2001){Brunetti}, {Setti}, {Feretti}, \&
  {Giovannini}}]{Brunetti_2001MNRAS.320..365B}
{Brunetti}, G., {Setti}, G., {Feretti}, L., \& {Giovannini}, G. 2001, \mnras,
  320, 365, \dodoi{10.1046/j.1365-8711.2001.03978.x}

\bibitem[{{Brunetti} \& {Vazza}(2020)}]{BV20}
{Brunetti}, G., \& {Vazza}, F. 2020, \prl, 124, 051101,
  \dodoi{10.1103/PhysRevLett.124.051101}

\bibitem[{{Brunetti} {et~al.}(2017){Brunetti}, {Zimmer}, \&
  {Zandanel}}]{Brunetti_2017MNRAS.472.1506B}
{Brunetti}, G., {Zimmer}, S., \& {Zandanel}, F. 2017, \mnras, 472, 1506,
  \dodoi{10.1093/mnras/stx2092}

\bibitem[{{Bryan} {et~al.}(2014){Bryan}, {Norman}, {O'Shea}, {Abel}, {Wise},
  {Turk}, {Reynolds}, {Collins}, {Wang}, {Skillman}, {Smith}, {Harkness},
  {Bordner}, {Kim}, {Kuhlen}, {Xu}, {Goldbaum}, {Hummels}, {Kritsuk}, {Tasker},
  {Skory}, {Simpson}, {Hahn}, {Oishi}, {So}, {Zhao}, {Cen}, {Li}, \& {Enzo
  Collaboration}}]{Enzo2014}
{Bryan}, G.~L., {Norman}, M.~L., {O'Shea}, B.~W., {et~al.} 2014, \apjs, 211,
  19, \dodoi{10.1088/0067-0049/211/2/19}

\bibitem[{{Burns} {et~al.}(1995){Burns}, {Roettiger}, {Pinkney}, {Perley},
  {Owen}, \& {Voges}}]{Burns_1995ApJ...446..583B}
{Burns}, J.~O., {Roettiger}, K., {Pinkney}, J., {et~al.} 1995, \apj, 446, 583,
  \dodoi{10.1086/175817}

\bibitem[{{Bustard} \& {Oh}(2022)}]{Bustard_Oh_2022}
{Bustard}, C., \& {Oh}, S.~P. 2022, \apj, 941, 65,
  \dodoi{10.3847/1538-4357/aca021}

\bibitem[{{Cassano} \& {Brunetti}(2005)}]{Cassano_Brunetti_2005}
{Cassano}, R., \& {Brunetti}, G. 2005, \mnras, 357, 1313,
  \dodoi{10.1111/j.1365-2966.2005.08747.x}

\bibitem[{{Cassano} {et~al.}(2016){Cassano}, {Brunetti}, {Giocoli}, \&
  {Ettori}}]{Cassano2016}
{Cassano}, R., {Brunetti}, G., {Giocoli}, C., \& {Ettori}, S. 2016, \aap, 593,
  A81, \dodoi{10.1051/0004-6361/201628414}

\bibitem[{{Cassano} {et~al.}(2010){Cassano}, {Brunetti}, {R{\"o}ttgering}, \&
  {Br{\"u}ggen}}]{Cassano_2010A&A...509A..68C}
{Cassano}, R., {Brunetti}, G., {R{\"o}ttgering}, H.~J.~A., \& {Br{\"u}ggen}, M.
  2010, \aap, 509, A68, \dodoi{10.1051/0004-6361/200913063}

\bibitem[{{Cassano} {et~al.}(2023){Cassano}, {Cuciti}, {Brunetti}, {Botteon},
  {Rossetti}, {Bruno}, {Simionescu}, {Gastaldello}, {van Weeren},
  {Br{\"u}ggen}, {Dallacasa}, {Zhang}, {Akamatsu}, {Bonafede}, {Di Gennaro},
  {Shimwell}, {de Gasperin}, {R{\"o}ttgering}, \& {Jones}}]{Cassano_2023}
{Cassano}, R., {Cuciti}, V., {Brunetti}, G., {et~al.} 2023, \aap, 672, A43,
  \dodoi{10.1051/0004-6361/202244876}

\bibitem[{{Cho} \& {Lazarian}(2006)}]{Cho_Lazarian_2006}
{Cho}, J., \& {Lazarian}, A. 2006, \apj, 638, 811, \dodoi{10.1086/498967}

\bibitem[{{Cho} {et~al.}(2009){Cho}, {Vishniac}, {Beresnyak}, {Lazarian}, \&
  {Ryu}}]{Cho_2009ApJ...693.1449C}
{Cho}, J., {Vishniac}, E.~T., {Beresnyak}, A., {Lazarian}, A., \& {Ryu}, D.
  2009, \apj, 693, 1449, \dodoi{10.1088/0004-637X/693/2/1449}

\bibitem[{{Cuciti} {et~al.}(2022){Cuciti}, {de Gasperin}, {Br{\"u}ggen},
  {Vazza}, {Brunetti}, {Shimwell}, {Edler}, {van Weeren}, {Botteon}, {Cassano},
  {Di Gennaro}, {Gastaldello}, {Drabent}, {R{\"o}ttgering}, \&
  {Tasse}}]{Cuciti_Megahalo}
{Cuciti}, V., {de Gasperin}, F., {Br{\"u}ggen}, M., {et~al.} 2022, \nat, 609,
  911, \dodoi{10.1038/s41586-022-05149-3}

\bibitem[{{del Valle} {et~al.}(2016){del Valle}, {de Gouveia Dal Pino}, \&
  {Kowal}}]{delValle_2016}
{del Valle}, M.~V., {de Gouveia Dal Pino}, E.~M., \& {Kowal}, G. 2016, \mnras,
  463, 4331, \dodoi{10.1093/mnras/stw2276}

\bibitem[{{Dennison}(1980)}]{Dennison_1980}
{Dennison}, B. 1980, \apjl, 239, L93, \dodoi{10.1086/183300}

\bibitem[{{Dom{\'\i}nguez-Fern{\'a}ndez}
  {et~al.}(2019){Dom{\'\i}nguez-Fern{\'a}ndez}, {Vazza}, {Br{\"u}ggen}, \&
  {Brunetti}}]{Dominguez-Fernandez_2019MNRAS.486..623D}
{Dom{\'\i}nguez-Fern{\'a}ndez}, P., {Vazza}, F., {Br{\"u}ggen}, M., \&
  {Brunetti}, G. 2019, \mnras, 486, 623, \dodoi{10.1093/mnras/stz877}

\bibitem[{{Eckert} {et~al.}(2019){Eckert}, {Ghirardini}, {Ettori}, {Rasia},
  {Biffi}, {Pointecouteau}, {Rossetti}, {Molendi}, {Vazza}, {Gastaldello},
  {Gaspari}, {De Grandi}, {Ghizzardi}, {Bourdin}, {Tchernin}, \&
  {Roncarelli}}]{Eckert_2019}
{Eckert}, D., {Ghirardini}, V., {Ettori}, S., {et~al.} 2019, \aap, 621, A40,
  \dodoi{10.1051/0004-6361/201833324}

\bibitem[{{En{\ss}lin} {et~al.}(2011){En{\ss}lin}, {Pfrommer}, {Miniati}, \&
  {Subramanian}}]{Ensslin_2011}
{En{\ss}lin}, T., {Pfrommer}, C., {Miniati}, F., \& {Subramanian}, K. 2011,
  \aap, 527, A99, \dodoi{10.1051/0004-6361/201015652}

\bibitem[{{Feretti} {et~al.}(1997){Feretti}, {Boehringer}, {Giovannini}, \&
  {Neumann}}]{Feretti_1997}
{Feretti}, L., {Boehringer}, H., {Giovannini}, G., \& {Neumann}, D. 1997, \aap,
  317, 432, \dodoi{10.48550/arXiv.astro-ph/9607027}

\bibitem[{{Fujita} {et~al.}(2007){Fujita}, {Kohri}, {Yamazaki}, \&
  {Kino}}]{Fujita_2007}
{Fujita}, Y., {Kohri}, K., {Yamazaki}, R., \& {Kino}, M. 2007, \apjl, 663, L61,
  \dodoi{10.1086/520337}

\bibitem[{{Fujita} {et~al.}(2003){Fujita}, {Takizawa}, \&
  {Sarazin}}]{Fujita_2003ApJ...584..190F}
{Fujita}, Y., {Takizawa}, M., \& {Sarazin}, C.~L. 2003, \apj, 584, 190,
  \dodoi{10.1086/345599}

\bibitem[{{Golovich} {et~al.}(2019){Golovich}, {Dawson}, {Wittman}, {Jee},
  {Benson}, {Lemaux}, {van Weeren}, {Andrade-Santos}, {Sobral}, {de Gasperin},
  {Br{\"u}ggen}, {Brada{\v{c}}}, {Finner}, \& {Peter}}]{Golovich_2019}
{Golovich}, N., {Dawson}, W.~A., {Wittman}, D.~M., {et~al.} 2019, \apjs, 240,
  39, \dodoi{10.3847/1538-4365/aaf88b}

\bibitem[{{Govoni} {et~al.}(2005){Govoni}, {Murgia}, {Feretti}, {Giovannini},
  {Dallacasa}, \& {Taylor}}]{Govoni_2005}
{Govoni}, F., {Murgia}, M., {Feretti}, L., {et~al.} 2005, \aap, 430, L5,
  \dodoi{10.1051/0004-6361:200400113}

\bibitem[{{Govoni} {et~al.}(2019){Govoni}, {Orr{\`u}}, {Bonafede}, {Iacobelli},
  {Paladino}, {Vazza}, {Murgia}, {Vacca}, {Giovannini}, {Feretti}, {Loi},
  {Bernardi}, {Ferrari}, {Pizzo}, {Gheller}, {Manti}, {Br{\"u}ggen},
  {Brunetti}, {Cassano}, {de Gasperin}, {En{\ss}lin}, {Hoeft}, {Horellou},
  {Junklewitz}, {R{\"o}ttgering}, {Scaife}, {Shimwell}, {van Weeren}, \&
  {Wise}}]{Govoni_2019}
{Govoni}, F., {Orr{\`u}}, E., {Bonafede}, A., {et~al.} 2019, Science, 364, 981,
  \dodoi{10.1126/science.aat7500}

\bibitem[{{Ignesti} {et~al.}(2020){Ignesti}, {Brunetti}, {Gitti}, \&
  {Giacintucci}}]{Ignesti_2020}
{Ignesti}, A., {Brunetti}, G., {Gitti}, M., \& {Giacintucci}, S. 2020, \aap,
  640, A37, \dodoi{10.1051/0004-6361/201937207}

\bibitem[{{Jaffe} \& {Rudnick}(1979)}]{Jaffe_1979}
{Jaffe}, W.~J., \& {Rudnick}, L. 1979, \apj, 233, 453, \dodoi{10.1086/157406}

\bibitem[{{Jeltema} \& {Profumo}(2011)}]{Jeltema_Profumo_2011}
{Jeltema}, T.~E., \& {Profumo}, S. 2011, ¥apj, 728, 53,
  \dodoi{10.1088/0004-637X/728/1/53}

\bibitem[{{Keshet} \& {Loeb}(2010)}]{Keshet_Loeb_2010}
{Keshet}, U., \& {Loeb}, A. 2010, \apj, 722, 737,
  \dodoi{10.1088/0004-637X/722/1/737}

\bibitem[{{Kowal} {et~al.}(2011){Kowal}, {de Gouveia Dal Pino}, \&
  {Lazarian}}]{Kowal_2011ApJ...735..102K}
{Kowal}, G., {de Gouveia Dal Pino}, E.~M., \& {Lazarian}, A. 2011, \apj, 735,
  102, \dodoi{10.1088/0004-637X/735/2/102}

\bibitem[{{Kowal} {et~al.}(2012){Kowal}, {de Gouveia Dal Pino}, \&
  {Lazarian}}]{Kowal_2012PhRvL}
---. 2012, \prl, 108, 241102, \dodoi{10.1103/PhysRevLett.108.241102}

\bibitem[{{Kravtsov} \& {Borgani}(2012)}]{Kravtsov_Borgani_2012}
{Kravtsov}, A.~V., \& {Borgani}, S. 2012, \araa, 50, 353,
  \dodoi{10.1146/annurev-astro-081811-125502}

\bibitem[{{Kunz} {et~al.}(2011){Kunz}, {Schekochihin}, {Cowley}, {Binney}, \&
  {Sanders}}]{Kunz_2011MNRAS.410.2446K}
{Kunz}, M.~W., {Schekochihin}, A.~A., {Cowley}, S.~C., {Binney}, J.~J., \&
  {Sanders}, J.~S. 2011, \mnras, 410, 2446,
  \dodoi{10.1111/j.1365-2966.2010.17621.x}

\bibitem[{{Lazarian} \& {Vishniac}(1999)}]{LV99}
{Lazarian}, A., \& {Vishniac}, E.~T. 1999, \apj, 517, 700,
  \dodoi{10.1086/307233}

\bibitem[{{Lazarian} \& {Xu}(2021)}]{Lazarian_Xu_2021}
{Lazarian}, A., \& {Xu}, S. 2021, \apj, 923, 53,
  \dodoi{10.3847/1538-4357/ac2de9}

\bibitem[{{Lazarian} \& {Xu}(2023)}]{Lazarian_Xu_2023}
---. 2023, arXiv e-prints, arXiv:2306.14973, \dodoi{10.48550/arXiv.2306.14973}

\bibitem[{{Lemoine}(2021)}]{Lemoine_2021}
{Lemoine}, M. 2021, \prd, 104, 063020, \dodoi{10.1103/PhysRevD.104.063020}

\bibitem[{{Lynn} {et~al.}(2014){Lynn}, {Quataert}, {Chandran}, \&
  {Parrish}}]{Lynn_2014}
{Lynn}, J.~W., {Quataert}, E., {Chandran}, B. D.~G., \& {Parrish}, I.~J. 2014,
  \apj, 791, 71, \dodoi{10.1088/0004-637X/791/1/71}

\bibitem[{{Miniati}(2015)}]{Miniati_2015ApJ...800...60M}
{Miniati}, F. 2015, \apj, 800, 60, \dodoi{10.1088/0004-637X/800/1/60}

\bibitem[{{Nelson} {et~al.}(2014){Nelson}, {Lau}, \& {Nagai}}]{Nelson_2014}
{Nelson}, K., {Lau}, E.~T., \& {Nagai}, D. 2014, \apj, 792, 25,
  \dodoi{10.1088/0004-637X/792/1/25}

\bibitem[{{Nishiwaki} \& {Asano}(2022)}]{Nishiwaki_2022ApJ}
{Nishiwaki}, K., \& {Asano}, K. 2022, \apj, 934, 182,
  \dodoi{10.3847/1538-4357/ac7d5e}

\bibitem[{{Nishiwaki} {et~al.}(2021){Nishiwaki}, {Asano}, \&
  {Murase}}]{Nishiwaki_2021ApJ...922..190N}
{Nishiwaki}, K., {Asano}, K., \& {Murase}, K. 2021, \apj, 922, 190,
  \dodoi{10.3847/1538-4357/ac1cdb}

\bibitem[{{Petrosian}(2001)}]{Petrosian_2001}
{Petrosian}, V. 2001, \apj, 557, 560, \dodoi{10.1086/321557}

\bibitem[{{Pfrommer} \& {En{\ss}lin}(2004)}]{Pfrommer_Ensslin_2004}
{Pfrommer}, C., \& {En{\ss}lin}, T.~A. 2004, \aap, 413, 17,
  \dodoi{10.1051/0004-6361:20031464}

\bibitem[{{Pinzke} {et~al.}(2017){Pinzke}, {Oh}, \& {Pfrommer}}]{Pinzke_2017}
{Pinzke}, A., {Oh}, S.~P., \& {Pfrommer}, C. 2017, \mnras, 465, 4800,
  \dodoi{10.1093/mnras/stw3024}

\bibitem[{{Pizzo} \& {de Bruyn}(2009)}]{Pizzo_2009}
{Pizzo}, R.~F., \& {de Bruyn}, A.~G. 2009, \aap, 507, 639,
  \dodoi{10.1051/0004-6361/200912465}

\bibitem[{{Pizzo} {et~al.}(2008){Pizzo}, {de Bruyn}, {Feretti}, \&
  {Govoni}}]{Pizzo_2008}
{Pizzo}, R.~F., {de Bruyn}, A.~G., {Feretti}, L., \& {Govoni}, F. 2008, \aap,
  481, L91, \dodoi{10.1051/0004-6361:20079304}

\bibitem[{{Porter} {et~al.}(2015){Porter}, {Jones}, \&
  {Ryu}}]{Porter_2015ApJ...810...93P}
{Porter}, D.~H., {Jones}, T.~W., \& {Ryu}, D. 2015, \apj, 810, 93,
  \dodoi{10.1088/0004-637X/810/2/93}

\bibitem[{{Press} \& {Schechter}(1974)}]{Press_Schechter_1974}
{Press}, W.~H., \& {Schechter}, P. 1974, \apj, 187, 425, \dodoi{10.1086/152650}

\bibitem[{{Ptuskin}(1988)}]{Ptuskin_1988}
{Ptuskin}, V.~S. 1988, Soviet Astronomy Letters, 14, 255

\bibitem[{{Schekochihin} {et~al.}(2009){Schekochihin}, {Cowley}, {Dorland},
  {Hammett}, {Howes}, {Quataert}, \&
  {Tatsuno}}]{Schekochihin_2009ApJS..182..310S}
{Schekochihin}, A.~A., {Cowley}, S.~C., {Dorland}, W., {et~al.} 2009, \apjs,
  182, 310, \dodoi{10.1088/0067-0049/182/1/310}

\bibitem[{{Schekochihin} {et~al.}(2005){Schekochihin}, {Cowley}, {Kulsrud},
  {Hammett}, \& {Sharma}}]{Schekochihin_2005ApJ...629..139S}
{Schekochihin}, A.~A., {Cowley}, S.~C., {Kulsrud}, R.~M., {Hammett}, G.~W., \&
  {Sharma}, P. 2005, \apj, 629, 139, \dodoi{10.1086/431202}

\bibitem[{{Schlickeiser} \& {Miller}(1998)}]{Schlickeiser_Miller_1998}
{Schlickeiser}, R., \& {Miller}, J.~A. 1998, \apj, 492, 352,
  \dodoi{10.1086/305023}

\bibitem[{{Steinwandel} {et~al.}(2022){Steinwandel}, {B{\"o}ss}, {Dolag}, \&
  {Lesch}}]{Steinwandel_2022}
{Steinwandel}, U.~P., {B{\"o}ss}, L.~M., {Dolag}, K., \& {Lesch}, H. 2022,
  \apj, 933, 131, \dodoi{10.3847/1538-4357/ac715c}

\bibitem[{{Steinwandel} {et~al.}(2023){Steinwandel}, {Dolag}, {B{\"o}ss}, \&
  {Marin-Gilabert}}]{Steinwandel_2023}
{Steinwandel}, U.~P., {Dolag}, K., {B{\"o}ss}, L., \& {Marin-Gilabert}, T.
  2023, arXiv e-prints, arXiv:2306.04692, \dodoi{10.48550/arXiv.2306.04692}

\bibitem[{{van Weeren} {et~al.}(2019){van Weeren}, {de Gasperin}, {Akamatsu},
  {Br{\"u}ggen}, {Feretti}, {Kang}, {Stroe}, \& {Zandanel}}]{vanWeeren_review}
{van Weeren}, R.~J., {de Gasperin}, F., {Akamatsu}, H., {et~al.} 2019, \ssr,
  215, 16, \dodoi{10.1007/s11214-019-0584-z}

\bibitem[{{Vazza} {et~al.}(2018){Vazza}, {Brunetti}, {Br{\"u}ggen}, \&
  {Bonafede}}]{vazza18dynamo}
{Vazza}, F., {Brunetti}, G., {Br{\"u}ggen}, M., \& {Bonafede}, A. 2018, \mnras,
  474, 1672, \dodoi{10.1093/mnras/stx2830}

\bibitem[{{Vazza} {et~al.}(2011){Vazza}, {Brunetti}, {Gheller}, {Brunino}, \&
  {Br{\"u}ggen}}]{Vazza_2011A&A...529A..17V}
{Vazza}, F., {Brunetti}, G., {Gheller}, C., {Brunino}, R., \& {Br{\"u}ggen}, M.
  2011, \aap, 529, A17, \dodoi{10.1051/0004-6361/201016015}

\bibitem[{{Vazza} {et~al.}(2017){Vazza}, {Jones}, {Br{\"u}ggen}, {Brunetti},
  {Gheller}, {Porter}, \& {Ryu}}]{Vazza_2017MNRAS.464..210V}
{Vazza}, F., {Jones}, T.~W., {Br{\"u}ggen}, M., {et~al.} 2017, \mnras, 464,
  210, \dodoi{10.1093/mnras/stw2351}

\bibitem[{{Vazza} {et~al.}(2016){Vazza}, {Wittor}, {Br{\"u}ggen}, \&
  {Gheller}}]{Vazza_2016}
{Vazza}, F., {Wittor}, D., {Br{\"u}ggen}, M., \& {Gheller}, C. 2016, Galaxies,
  4, 60, \dodoi{10.3390/galaxies4040060}

\bibitem[{{Vazza} {et~al.}(2023){Vazza}, {Wittor}, {Di Federico},
  {Br{\"u}ggen}, {Brienza}, {Brunetti}, {Brighenti}, \&
  {Pasini}}]{2023A&A...669A..50V}
{Vazza}, F., {Wittor}, D., {Di Federico}, L., {et~al.} 2023, \aap, 669, A50,
  \dodoi{10.1051/0004-6361/202243753}

\bibitem[{{Xu} \& {Lazarian}(2016)}]{Xu_Lazarian_2016}
{Xu}, S., \& {Lazarian}, A. 2016, \apj, 833, 215,
  \dodoi{10.3847/1538-4357/833/2/215}

\bibitem[{{Yuan} {et~al.}(2003){Yuan}, {Zhou}, \& {Jiang}}]{Yuan_2003}
{Yuan}, Q., {Zhou}, X., \& {Jiang}, Z. 2003, \apjs, 149, 53,
  \dodoi{10.1086/380936}

\bibitem[{{ZuHone} {et~al.}(2015){ZuHone}, {Brunetti}, {Giacintucci}, \&
  {Markevitch}}]{ZuHone_2015}
{ZuHone}, J.~A., {Brunetti}, G., {Giacintucci}, S., \& {Markevitch}, M. 2015,
  \apj, 801, 146, \dodoi{10.1088/0004-637X/801/2/146}

\bibitem[{{ZuHone} {et~al.}(2011){ZuHone}, {Markevitch}, \&
  {Lee}}]{ZuHone_2011ApJ...743...16Z}
{ZuHone}, J.~A., {Markevitch}, M., \& {Lee}, D. 2011, \apj, 743, 16,
  \dodoi{10.1088/0004-637X/743/1/16}

\end{thebibliography}
\bibliographystyle{aasjournal}

\appendix

\section{Cosmic-ray protons and secondary electrons}\label{app:secondary}

In addition to the turbulent reacceleration, the injection of secondary leptons from the the decay of pions produced by $pp$ collision between CRp and thermal protons is often invoked for the mechanism for the diffuse radio emission in galaxy clusters \citep[e.g.,][]{Dennison_1980,Blasi_1999APh....12..169B,Keshet_Loeb_2010,Ensslin_2011}.
The limits on the gamma ray emission from ICM \citep[e.g.,][]{Jeltema_Profumo_2011,Ackermann_2014,Ackermann_2016} suggest that the contribution of secondary particle production through $pp$ collision to the observed emission is sub-dominant \citep[][]{Brunetti_2017MNRAS.472.1506B,Adam_2021}.
However, the secondary leptons can contribute to the diffuse radio emission by providing the seed population for the reacceleration \citep[e.g.,][]{Brunetti_2017MNRAS.472.1506B,Pinzke_2017,Nishiwaki_2021ApJ...922..190N}.
Also, the mini halos in the dense core of cool-core clusters can originate from the injection of secondaries \citep[e.g.,][]{Pfrommer_Ensslin_2004,Fujita_2007,ZuHone_2015,Ignesti_2020}. 
In this section, we explore the contribution of the secondary electrons to the diffuse emission found in cluster outskirts. 
\par



We calculate the injection of secondary CRe for a given distribution of CRp, using the procedure of \citet{Nishiwaki_2021ApJ...922..190N}.
We assume a single power-law spectrum of CRp, $N_p\propto p^{-\alpha_p}$, with $\alpha_p = 3.2$ to fit the radio synchrotron spectrum with $\alpha_{\rm syn}\approx1.6$.
We set $p/(m_{\rm p}c) = 1$ for the minimum momentum of CRp.
The energy density of CRp $\epsilon_{\rm CRp}$ is treated as a free parameter.
For simplicity, we neglect the re-acceleration of both CRe and CRp.
We adopt the one-zone approximation explained in Sect.~\ref{subsec:onezone}.
\par





\begin{figure}
	\includegraphics[width=\columnwidth]{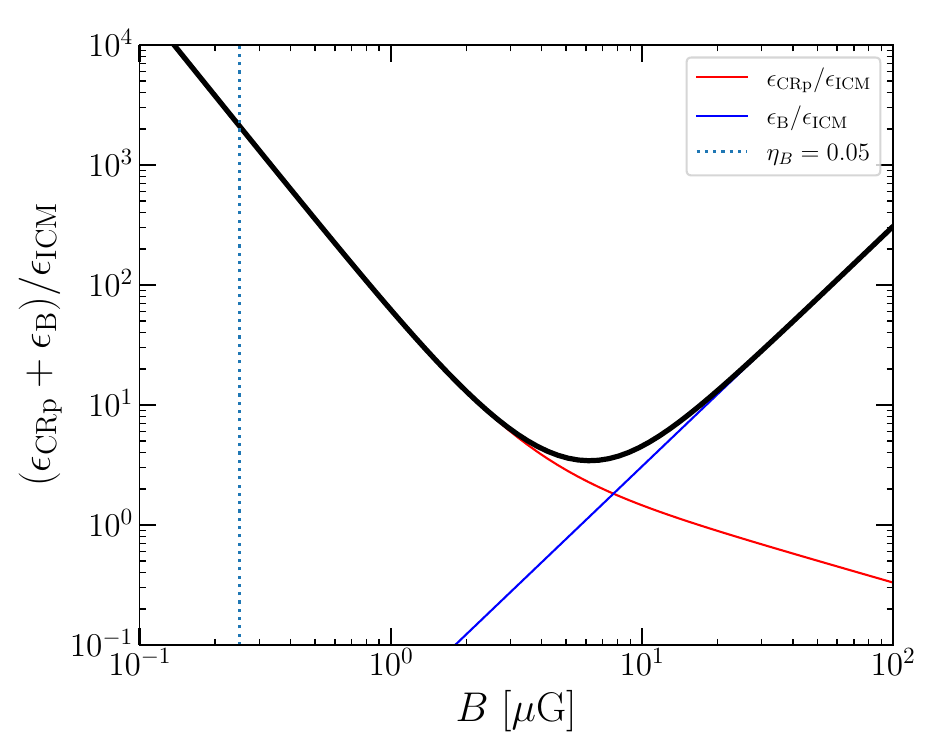}
    \caption{
    Energy densities of CRp and magnetic field required in the pure hadronic model without reacceleration as a function of the magnetic field.
    The red and blue lines show $\epsilon_{\rm CRp}$ and $\epsilon_{\rm B}$, respectively.
    The thick black line shows the sum of those two.
    The energy densities are normalized by that of the ICM in the simulated box ($\epsilon_{\rm ICM} = 1.3 \times 10^{-13}~{\rm erg/cm^3}$).
    The spectral index of CRp is assumed to be $\alpha_{p} = 3.2$.
    The vertical dotted line shows the magnetic field calculated with $\eta_{B} = 0.05$ (Sect.~\ref{sec:theory}).}
    \label{fig:pure_hadronic}
\end{figure}
In Fig.~\ref{fig:pure_hadronic}, we show the energy densities of CRp and the magnetic field required to explain the observed brightness at 49 MHz in the pure hadronic model without reacceleration.
We find that $\epsilon_{\rm CRp}$ should be as large as $\epsilon_{\rm CRp} \approx 10^3\epsilon_{\rm ICM}$ to explain the observed brightness when we adopt the same magnetic field as Sect.~\ref{subsec:onezone}, i.e., $B = 0.24~\mu{\rm G}$.
The energy density of secondary CRe is negligible compared to that of CRp.
The synchrotron brightness scales as $S_{\rm syn}\propto K_eB^{(\delta+1)/2}$, where $K_e\propto \epsilon_{\rm CRp}/(B^2+B^2_{\rm CMB}(z))$ and $\delta\approx\alpha_{p}+1$ are the normalization and the power-law index of secondary CRe spectrum, respectively \citep[e.g.,][]{Brunetti_Jones_review}.
We find that the minimum value of the non-thermal energy density $(\epsilon_{\rm CRp} + \epsilon_{B})$ is 3 times larger than $\epsilon_{\rm ICM}$, and it appears when the magnetic field is as large as $B = 6~\mu{\rm G}$.
Thus, the classical ``pure-hadronic" model, which does not include any reacceleration, is incompatible with the observed mega halo.

Note that the above discussion does not exclude the possibility that the diffuse emission is produced by the reacceleration of the secondaries injected through the $pp$ collision.
A comprehensive study of this ``secondary re-acceleration" model should incorporate the re-acceleration of CRp, which is out of the scope of this work.
It is worth noting that turbulence might be significantly damped by the back reaction of the reacceleration of CRp when $\epsilon_{\rm CRp} \gtrsim \epsilon_{\rm turb}$ \citep[e.g.,][]{Brunetti_Lazarian_2007}.
Such back reaction is not considered in the derivation of Eq.~(\ref{eq:Dpp}).


\section{dependence on the initial condition}\label{app:initial}

In Sect.~\ref{sec:FP}, we assumed that the initial spectrum before the onset of the reacceleration, $N_e(p,0)$, is determined by the radiative and the Coulomb cooling.
The spectrum has a peak around $p_{\rm min}\sim10$ as shown in Fig.~\ref{fig:spectrum}.
However, it is also possible that the ``initial spectrum" is affected by the turbulent reacceleration working before the time of the snapshot.
Recent simulation by \citet{Beduzzi_2023arXiv230603764B} observed that the radio-emitting CRe in the ICM experience multiple episodes of reacceleration.
In such a case, the peak of the seed CRe spectrum should appear at a larger momentum $p_{\rm min} \sim 10^2-10^3$.
In this section, we discuss how the results in Sect.~\ref{subsec:cell} are modified when we adopt a different initial spectrum.
We adopt the same method as Sect.~\ref{subsec:cell}, taking into account the LOS integration.
We fix $\eta_B = 0.05$ also in this section.

\par


To consider the reacceleration before the epoch of the snapshot, we assume the initial spectrum for the calculation with $p_{\rm min} = 10^3$.
The shape of the initial spectrum is assumed to be the same in all cells.
The initial energy density of the CRe follows $\epsilon_{\rm CRe}(t = 0) \propto \epsilon_{\rm ICM}$, as in Sect.~\ref{subsec:cell}.

\par

The difference of the initial spectrum is not important when $T_{\rm dur}$ (calculation time of the FP equation) is very long and the steady state due to the balance between the cooling and reacceleration is achieved in most of the cells.
In Sect.~\ref{sec:FP}, we find that the steady state is achieved at $T_{\rm dur} \geq 3$ Gyr when $t_{\rm acc}\approx0.5$ Gyr.

In this Section, we limit $T_{\rm dur}$ by $\approx 2t_{\rm eddy}$, where $t_{\rm eddy}$ is the eddy turnover time measured in each cell. For the cells with $2t_{\rm eddy}\gtrsim 1~{\rm Gyr}$, we terminate the calculation at 1 Gyr.
The mean value of $t_{\rm eddy}$ in the simulated box is $\approx0.5$ Gyr, so $T_{\rm dur} = 2t_{\rm eddy}$ is shorter than 3 Gyr in most of the cells.


\par


\begin{figure*}
    \plottwo{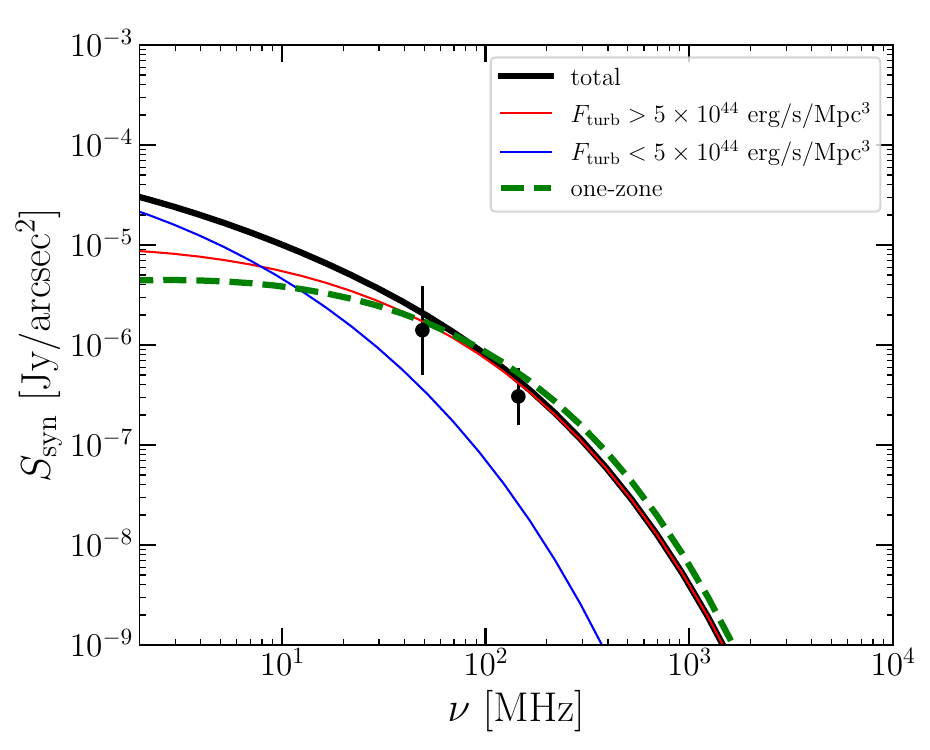}{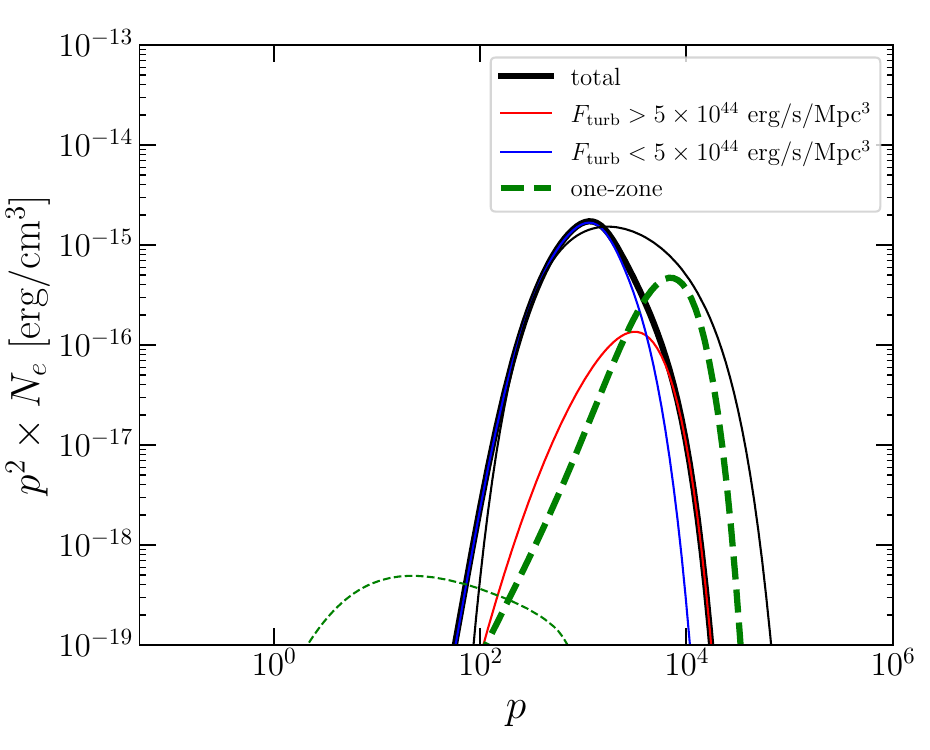}
    \caption{Same as Fig.~\ref{fig:spectrum}, but the minimum momentum of the initial spectrum is assumed to be $p_{\rm min}  = 10^3$ due to the reacceleration before the epoch of the snapshot. The duration of the FP calculation is limited by $2t_{\rm eddy}$ and $\psi = 0.5$ is adopted.}
    \label{fig:spec_app}
\end{figure*}

With the above conditions, we find that the typical spectral index,$\alpha_{\rm syn}\approx1.6$, can be reproduced with $\psi = 0.5$.
Fig.~\ref{fig:spec_app} (right) shows the CRe spectra before and after the calculation.
Summing up the CRe spectra over cells included in one beam ($\sim10^3$ cells), we find that $p_{\rm min}$ after the calculation does not change much from the initial value, $p_{\rm min} = 10^3$.
This is consistent with the assumption that $p_{\rm min} = 10^3$ is caused by the turbulent acceleration prior to the epoch of the snapshot.
We calculate the acceleration efficiency using Eq.~(\ref{eq:eta_acc_beam}) and find a typical value of $\langle\eta_{\rm acc}\rangle \approx 1.7\%$.

Although we assumed a shorter $T_{\rm dur}$ than that in Sect.~\ref{subsec:cell}, the contribution of the cells with smaller turbulent energy ($F_{\rm turb}<5\times10^{44}$ erg/s/Mpc$^3$) increases, and almost 60\% of the area of emission region can be covered at the LOFAR sensitivity.

\par

If we adopt the initial spectrum with $p_{\rm min}\sim10$ and calculate reacceleration for $T_{\rm dur} = 2t_{\rm eddy}$, the number of CRe with $p\sim10^4$ at the final state becomes smaller than the $p_{\rm min}\sim10^3$ case and that model cannot explain the observed brightness unless $\eta_{\rm acc}\gtrsim10\%$.

\end{document}